\documentclass[a4paper,11pt]{article}

\pdfoutput=1

\usepackage{jcappub_ver}
\usepackage{amsmath}
\usepackage{graphicx}

\usepackage{physics}
\usepackage{pgfplots}

\usepackage{verbatim}

\usepackage[english]{babel}
\usepackage[utf8]{inputenc}
\usepackage{pdflscape}
\usepackage{enumerate}
\usepackage{amsbsy}
\usepackage{amsmath} 
\usepackage{graphics}
\usepackage{mathrsfs}
\usepackage{wrapfig}
\usepackage{mathtools}

\usepackage{amsfonts}
\usepackage{pstricks}
\usepackage{color}
\usepackage{setspace}
\usepackage{tensor}
\usepackage[ruled,vlined]{algorithm2e}
\usepackage{array}
\usepackage{booktabs}
\usepackage{footmisc}
\usepackage{changepage}
\usepackage[normalem]{ulem}
\usepackage{float}

\newcommand{\bea}{\begin{eqnarray}} \newcommand{\eea}{\end{eqnarray}}
\newcommand{\el}{\nonumber \\}
\newcommand{\re}[1]{(\ref{#1})}

\newcommand{\pat}{\partial}
\renewcommand{\sec}[1]{section \ref{#1}}
\newcommand{\fig}[1]{figure \ref{#1}}
\newcommand{\tab}[1]{table \ref{#1}}

\newcommand{\para}{\paragraph}

\renewcommand*{\vec}[1]{\mathbf{#1}}

\renewcommand{\a}{\alpha}
\renewcommand{\b}{\beta}
\renewcommand{\c}{\gamma}
\renewcommand{\d}{\delta}

\newcommand{\GN}{G_{\mathrm{N}}}
\newcommand{\ha}{\frac{1}{2}}
\newcommand{\rmd}{\mathrm{d}}

\newcommand{\ie}{i.e.\ }
\newcommand{\eg}{e.g.\ }

\newcommand{\av}[1]{\langle{#1}\rangle}

\newcommand{\phil}{\bar\phi}

\newcommand{\rmax}{r_{\text{max}}}
\newcommand{\kmax}{k_{\text{max}}}

\newcommand{\C}{\mathcal{C}}
\newcommand{\avC}{\bar{\mathcal{C}}}
\newcommand{\Cmax}{\C_{\text{max}}}
\newcommand{\avCmax}{\avC_{\text{max}}}
\newcommand{\zetamax}{\zeta_{\text{max}}}
\newcommand{\zetath}{\zeta_{\text{th}}}
\newcommand{\Cth}{\C_\text{th}}
\newcommand{\avCth}{\avC_\text{th}}

\newcommand{\kc}{k_{\sigma}}
\newcommand{\kpeak}{k_\text{peak}}

\newcommand{\Pzeta}{\mathcal{P}_\zeta}

\newcommand{\bk}{\mathbf{k}}
\newcommand{\bp}{\mathbf{p}}
\newcommand{\bx}{\mathbf{x}}
\newcommand{\hxi}{{\hat{\xi}}}
\newcommand{\tk}{{\tilde{k}}}
\newcommand{\kini}{k_\text{ini}}
\newcommand{\kend}{k_\text{end}}
\newcommand{\thr}{\mathrm{th}}
\newcommand{\kold}{k_\text{USR}}

\newcommand{\paperI}{$Paper\,I$}

\begin{document}

\title{Effect of stochastic kicks on primordial black hole abundance and mass via the compaction function}

\author[a]{Sami Raatikainen}
\author[a, b]{\hspace*{-2mm}, Syksy R\"{a}s\"{a}nen}
\author[c, d]{\hspace*{-2mm}, and Eemeli Tomberg}

\affiliation[a]{University of Helsinki, Helsinki Institute of Physics,\\ P.O. Box 64, FIN-00014 University of Helsinki, Finland}

\affiliation[b]{University of Helsinki, Department of Physics,\\ P.O. Box 64, FIN-00014 University of Helsinki, Finland}

\affiliation[c]{Consortium for Fundamental Physics, Physics Department, \\ Lancaster University, Lancaster LA1 4YB, United Kingdom}

\affiliation[d]{Cosmology, Universe and Relativity at Louvain (CURL), Institute of Mathematics and Physics, University of Louvain, 2 Chemin du Cyclotron, 1348 Louvain-la-Neuve, Belgium}

\emailAdd{sami.raatikainen@iki.fi}
\emailAdd{syksy.rasanen@iki.fi}
\emailAdd{eemeli.tomberg@uclouvain.be}

\begin{flushleft}
	\hfill		 HIP-2025-28/TH \\
\end{flushleft}

\abstract{We study stochastic effects in viable ultra-slow-roll inflation models that produce primordial black holes. We consider asteroid, solar, and supermassive black hole seed masses. In each case, we simulate $10^8$ patches of the universe that may collapse into PBHs. In every patch, we follow $4\times10^4$ momentum shells to construct its spherically symmetric profile from first principles, without introducing a window function. We include the effects of critical collapse and the radiation era transfer function. The resulting compaction function profiles are very spiky due to stochastic kicks. This can enhance the PBH abundance by up to 36 orders of magnitude, depending on the mass range and collapse criterion. The PBH mass function shifts to higher masses and widens significantly. These changes may have a large effect on observational constraints of PBHs and make it possible to generate PBHs with a smaller amplitude of the power spectrum. However, convergence issues for the mass function remain. The results call for redoing collapse simulations to determine the collapse criterion for spiky profiles.
}

\maketitle
  
\setcounter{tocdepth}{2}

\setcounter{secnumdepth}{3}

\section{Introduction} \label{sec:intro}

Primordial black holes (PBHs) were first suggested as dark matter by Stephen Hawking in 1971 \cite{Hawking:1971ei} (PBHs were introduced in 1967 by Zel'dovich and Novikov \cite{Zeldovich:1967lct}; see \cite{Carr:2024nlv} for a brief history). They can constitute all of the dark matter if their masses are in the asteroid mass range or, if Hawking radiation stops or is damped at some point, for smaller masses \cite{Carr:2020gox, Green:2020jor, Green:2024bam}. PBHs could also act as the seeds of supermassive black holes at the centres of galaxies or even early galaxies themselves: it is not clear how those objects can have become very massive very rapidly, a problem heightened by recent observations by the James Webb Space Telescope \cite{Inayoshi:2019fun, Maiolino:2023zdu, Zhang:2025asq, Matteri:2025vnv, Prole:2025snf}. Generating the PBHs requires large fluctuations on small scales. It seems a natural possibility that these are produced in the same process as the small fluctuations on large scales seen in the cosmic microwave background (CMB) and large-scale structure. The most successful scenario for this is cosmic inflation \cite{Starobinsky:1979ty, Starobinsky:1980te, Kazanas:1980tx, Guth:1980zm, Sato:1980yn, Mukhanov:1981xt, Linde:1981mu, Albrecht:1982wi, Hawking:1981fz, Chibisov:1982nx, Hawking:1982cz, Guth:1982ec, Starobinsky:1982ee, Sasaki:1986hm, Mukhanov:1988jd}, suggested as the origin of PBHs in \cite{Dolgov:1992pu, Ivanov:1994pa}.

Generating the large inhomogeneities needed for PBHs requires large quantum fluctuations during inflation, which can have a significant effect on the mean evolution. Stochastic inflation is a formalism to treat the interaction between the mean and the perturbations. The field is split into long-wavelength modes described by classical non-linear theory with a stochastic noise term given by the short-wavelength modes, which are treated in linear quantum theory \cite{Starobinsky:1986fx, Morikawa:1989xz, Habib:1992ci, Mijic:1994vv, Starobinsky:1994bd} (see \cite{Vennin:2024yzl} for a brief review). In single-field inflation, the comoving curvature perturbation is inversely proportional to the time derivative of the background field, so the background field has to slow down to produce large perturbations. A particularly well-studied case is ultra-slow-roll (USR) inflation, where the gradient of the potential (\ie classical drift) is either small in comparison to the Hubble friction or flips sign and makes the field momentarily climb uphill \cite{Ivanov:1994pa, Faraoni:2000vg, Leach:2001zf, Kinney:2005vj, Martin:2012pe, Garcia-Bellido:2017mdw, Ezquiaga:2017fvi, Kannike:2017bxn, Germani:2017bcs, Motohashi:2017kbs, Dimopoulos:2017ged, Gong:2017qlj, Ballesteros:2017fsr, Hertzberg:2017dkh, Pattison:2018, Biagetti:2018pjj, Ezquiaga:2018gbw, Rasanen:2018fom, Karam:2022nym}. When the field moves slowly, stochastic kicks become important and produce an exponential tail for the distribution of the perturbation amplitude. It can dominate over the linear theory Gaussian tail and enhance PBH abundance by several orders of magnitude \cite{Pattison:2017mbe, Ezquiaga:2019ftu, Figueroa:2020jkf, Pattison:2021oen, Tomberg:2021xxv, Figueroa:2021zah, Cai:2022erk, Tomberg:2022mkt, Gow:2022jfb, Tomberg:2023kli, Launay:2024qsm, Animali:2024jiz, Vennin:2024yzl, Inui:2024sce, Sharma:2024fbr} (see \cite{Cruces:2018cvq, Cruces:2021iwq, Cruces:2022dom, Cruces:2024pni} for arguments to the contrary, and also \cite{Artigas:2023kyo, Artigas:2025nbm}). (As the perturbations are large, classical non-linearities can also lead to a heavy tail \cite{Biagetti:2021eep,  Kitajima:2021fpq, Hooshangi:2021ubn, Cai:2021zsp, DeLuca:2022rfz, Ferrante:2022mui, Gow:2022jfb, Pi:2022ysn, Hooshangi:2023kss, Wang:2024xdl, Inui:2024sce}.) After inflation, rare patches of space with large inhomogeneities will collapse to form PBHs \cite{Carr:1975qj, Nadezhin:1978, Bicknell:1979}, and the observed abundance and mass distribution depends on the initial number density and structure of these patches.

In our letter \cite{Raatikainen:2023bzk} -- which we refer to as \paperI\ -- we showed for the first time that the stochastic kicks that lead to the exponential tail for the probability distribution of the patches also make the radial profiles of the individual patches very choppy, which can facilitate collapse into PBHs and thus further increase their abundance. We calculated the abundance and mass distribution of asteroid-mass PBHs by simulating $10^9$ patch realisations, finding the radial profile of the compaction function for each, and assuming that PBHs form if the compaction function or its average exceeds a threshold determined from simulations of smooth profiles. (It was argued in \cite{Kehagias:2024kgk} that only average profiles can be calculated from stochastic inflation; this is not the case.)

We now extend the calculation to PBHs with masses around $M_\odot$ or $10^3 M_\odot$, the latter being suitable seeds for supermassive black holes. We also consider the asteroid case in more detail, and include the effects of critical collapse and the transfer function on the mass function. In \sec{sec:form} we go over the formalism and our simulation procedure. Some material is repeated from \paperI\ and our earlier works \cite{Figueroa:2020jkf, Figueroa:2021zah} to have a self-contained presentation. In \sec{sec:res} we present our results, and in \sec{sec:disc} we discuss some of their implications. In \sec{sec:conc} we summarise our findings and highlight open questions in building a reliable path from the inflaton potential to the PBH abundance. Details of numerical resolution and convergence are discussed in appendix \ref{app:res}.

\section{Formalism for calculating PBH abundance} \label{sec:form}

\subsection{Stochastic inflation and the $\Delta N$ formalism} \label{sec:stoc}

Inflationary calculations mostly rely on a split into background and perturbations, and the range of numerical relativity and lattice studies remains limited. As inflation stretches the wavelength of nearly massless modes, they become part of the background for modes with smaller wavelengths, so there is a flow from the perturbations to the background. As the initial amplitude of the modes is determined by a quantum process, the background evolution becomes subject to stochastic noise. In this setup, the background is the system and the small-scale modes are the environment that it evolves in \cite{Salopek:1992qy, Sasaki:1995aw, Sasaki:1998ug, Wands:2000dp, Lyth:2004gb}. At the same time, the background determines the evolution of the small-scale modes, leading to coupled equations that have only recently been solved consistently at every timestep for each realisation \cite{Figueroa:2020jkf, Figueroa:2021zah} (see \cite{Levasseur:2013ffa, Levasseur:2013tja, Levasseur:2014ska} for earlier recursive calculations). The stochastic kicks can be expected to be important when inflation lasts long \cite{Tsamis:2005hd, Woodard:2005cv, Ando:2020fjm, Woodard:2025cez} or when the kicks are large.

In the $\Delta N$ formalism, the curvature perturbation is given by the difference in the duration of inflation quantified by the number of e-folds $N$. The term $\Delta N$ formalism is used in the literature in three different ways, which can be considered successive levels of approximation. The first is the original application, where different patches of the universe are taken to evolve like unperturbed Friedmann--Lema\^itre--Robertson--Walker universes, their differences contained in the local value of the time coordinate \cite{Salopek:1992qy, Sasaki:1995aw, Sasaki:1998ug, Wands:2000dp, Lyth:2004gb}. The second is the stochastic $\Delta N$ formalism, where a patch's evolution is affected by small-scale modes within it, so each patch does not follow the FLRW equations, but $\Delta N$ is still the only quantity of interest \cite{Finelli:2008zg, Finelli:2010, Fujita:2013cna, Vennin:2015hra, Pattison:2019hef, Artigas:2024ajh}. The third case is when the patches do not necessarily evolve like FLRW universes and more information than simply the overall expansion $\Delta N$ is followed for each patch, such as shear to capture gravitational waves. This is sometimes called the extended or generalised $\Delta N$ formalism, although the term separate universe approximation might be more appropriate \cite{Naruko:2012fe, Tanaka:2021dww, Tanaka:2023gul, Artigas:2024ajh, Tanaka:2024mzw}.

Our application of the stochastic formalism is close to the second approximation, but although we consider only $\Delta N$, we keep many $\Delta N$ values from the same patch to resolve details of its spatial structure beyond the overall expansion. In the stochastic formalism, the field does not, in general, evolve along the classical FLRW trajectory in phase space, but in our case this is a good approximation, since the classical background solution is an attractor and the modes of interest become highly squeezed along this attractor. Nevertheless, the motion back and forth along the attractor trajectory is stochastic, not deterministic \cite{Tomberg:2022mkt}. For rapid transitions between slow-roll and USR, the $\Delta N$ formalism may fail as gradients become important; we do not take this into account \cite{Naruko:2012fe, Jackson:2023obv, Domenech:2023dxx, Artigas:2024ajh, Raveendran:2025pnz, Prokopec:2025uvz, Briaud:2025ayt}.

We consider single-field inflation with a minimally coupled scalar field with a canonical kinetic term (we use units such that the reduced Planck mass is unity),
\bea \label{action}
  S = \int \rmd^4 x \sqrt{-g} \left[ \ha R - \ha g^{\a\b} \pat_\a \phi \pat_\b \phi - V(\phi) \right] \ ,
\eea
where $R$ is the Ricci scalar and $\phi$ is the inflaton.

We decompose the inflaton into long and short wavelength parts as
\bea \label{dec}
  \phi(t,\bx) &=& \int \frac{\rmd^3 k}{(2\pi)^\frac{3}{2}} \phi_\vec{k}(t) e^{-i\vec{k} \cdot \bx} \el
  &=& \int \frac{\rmd^3 k}{(2\pi)^\frac{3}{2}} \left[ 1 - W\left(\frac{k}{\sigma a H}\right) \right] \phi_\vec{k}(t) e^{-i\vec{k} \cdot \bx} + \int \frac{\rmd^3 k}{(2\pi)^\frac{3}{2}} W\left(\frac{k}{\sigma a H}\right) \phi_\vec{k}(t) e^{-i\vec{k} \cdot \bx} \el
  &\equiv& \phil(t,\bx) + \delta\phi(t,\bx) \ ,
\eea
where $k\equiv|\vec{k}|$, $a(t)$ is the scale factor, $H\equiv\frac{\rmd \ln a}{\rmd t}$ is the Hubble parameter, and $W\left(\frac{k}{\sigma a H}\right)$ is a window function, which we take to be the step function: $W\left(\frac{k}{\sigma a H}\right)=\theta(k-\sigma a H)$ \cite{Starobinsky:1986fx, Morikawa:1989xz, Habib:1992ci, Mijic:1994vv, Starobinsky:1994bd}. The parameter $\sigma$ controls the separation scale $\kc(t) \equiv \sigma aH$: modes with $k \ll \kc$ are considered long wavelength and modes with $k \gg \kc$ short wavelength. In the inflationary models we consider, $H$ changes little during the relevant period, so we take $H=H_0$ in the window function, where $H_0$ is the initial value of the Hubble parameter. All our simulations start from scales that exit the Hubble radius from about 52 e-folds before the end of inflation. To leading order in the gradient expansion, the long-wavelength part $\bar\phi$ evolves according to the following stochastic equations (see \eg \cite{Pattison:2019hef, Figueroa:2021zah, Tomberg:2022mkt}):
\bea \label{bg_eom}
  \dot{\bar\phi} = \bar\pi + \xi_\phi \ , \quad
  \dot{\bar\pi} = -\qty(3 - \frac{1}{2}\bar\pi^2)\qty( \bar\pi + \frac{V_{,\bar\phi}(\bar\phi)}{V(\bar\phi)} ) + \xi_\pi \ ,
\eea
where dot denotes derivative with respect to the number of e-folds $N$, $\bar\pi$ is the mean field momentum, and $\xi_\phi$ and $\xi_\pi$ are the stochastic noise for the field and the momentum, respectively. The small wavelength modes are instead treated in linear theory, by solving the Mukhanov--Sasaki equation, so their spectrum is Gaussian, and the noise correlators are
\bea \label{xi}
  \av{\xi_X(N)\xi_Y(\tilde{N})} &=& \mathcal P_{X Y}(N,\kc)\delta(N-\tilde{N}) \ ,
\eea
where $\mathcal P_{X Y}(N, k) \equiv \frac{k^3}{2\pi^2} X_k(N) Y_k(N)^*$, and $X, Y\in\{\d\phi, \delta\pi\}$. As mentioned above, on the attractor the perturbations are squeezed and the field noise and the momentum noise are correlated, so the field moves back and forth only along the deterministic non-stochastic background trajectory in phase space, and there is just one independent noise term. The noise is white because we have chosen the window function in \re{dec} to be the step function. This approximation, which employs a sharp distinction between the background and the perturbations, offers considerable technical simplification \cite{Morikawa:1989xz, Casini:1998wr, Matarrese:2003ye, Liguori:2004fa, Levasseur:2014ska, Grain:2017dqa, Mahbub:2022osb}. The power spectrum of the curvature perturbation is\footnote{To be precise, this is the power spectrum of the curvature perturbation in the comoving gauge, $\mathcal{R}$. The quantity $\zeta$ is the curvature perturbation in the uniform energy density gauge. In the deep super-Hubble regime at times sufficiently long after the end of USR, the case of interest for our computations, we have $\zeta \simeq \mathcal{R}$ (possibly with a minus sign, depending on the convention) \cite{Malik:2008im}.} $\mathcal P_\zeta=\mathcal P_{\d\phi \d\phi}/\dot{\bar\phi}^2$. The field slows down in USR and speeds up after it, which produces a peak in the power spectrum; we denote the location of the maximum by $\kpeak$.

We consider the inflation models introduced in \cite{Figueroa:2020jkf, Figueroa:2021zah}, based on a Higgs inflation potential that includes loop corrections \cite{Rasanen:2018fom}, and adjusted by hand to make the CMB region agree with observations \cite{Akrami:2018odb}\footnote{In the solar case, the tensor-to-scalar ratio is slightly above the current upper limit \cite{BICEP:2021xfz}. In all three cases, the spectral index is in tension with the recent combination of ACT CMB data and DESI large-scale structure data \cite{ACT:2025fju}; see also \cite{Ferreira:2025lrd}. These issues would be easy to correct, with minimal change to our results.}. Loop corrections create a local hilltop in the potential, and USR takes place as the field climbs up the hill. On the other side of the hill, the field enters a dual phase of attractor constant-roll inflation. The hilltop location determines the PBH mass. We study three cases. The hilltop is tuned so that if stochastic effects were neglected and the abundance calculated with the threshold $\zetath$ on the comoving curvature perturbation, then in the first two cases PBHs with either asteroid or solar mass would constitute practically all of the dark matter, and in the third case they would provide a suitable number density of $\sim10^3 M_\odot$ seeds for supermassive black holes\footnote{Assuming efficient accretion: see however \cite{Prole:2025snf}.}.

As in \paperI, we do not solve the stochastic equations \re{bg_eom} and \re{xi} directly at every timestep, unlike in our initial work \cite{Figueroa:2020jkf, Figueroa:2021zah}. Instead, we take advantage of approximations suited to the USR phase and the following dual constant-roll phase. In typical models, stochasticity is strongest when the field is in the constant-roll attractor, where the equations have the following analytical solution in terms of discrete $k$ modes \cite{Tomberg:2023kli}:
\bea \label{phi_solution}
    \bar\phi(N) = \phi_0 e^{\frac{\epsilon_2}{2}N} \qty(1 - \frac{\epsilon_2}{2}X_{<\kc(N)}) \ ,
\eea
where $\phi_0$ is an integration constant and $\epsilon_2 \equiv \frac{\rmd \ln{}(-\dot H/H)}{\rmd N}$ is the second slow-roll parameter during the constant-roll phase when it is approximately constant. The variable $X_{<k}$ is a sum over Gaussian random variables $\hxi_k$,
\bea \label{X_solution}
    X_{<k} \equiv -\!\sum_{\tk=\kini}^{k} \sqrt{\Pzeta(\tk) \, \rmd \ln \tk} \, \hxi_\tk \ ,
\eea
where $\kini$ is the wavenumber of the first \ie longest wavelength mode considered, and $\hxi_k$ is a discrete version of the Gaussian noise with zero mean and unit variance, $\av{\hxi_k\hxi_\tk} = \delta_{k\tk}$. The results are not sensitive to the precise choice of $\kini$, as long as $\kini\ll\kpeak$, so that the amplitude of the neglected modes is small. The power spectrum $\Pzeta(k)$ is pre-solved from the Mukhanov--Sasaki equation using deterministic evolution for the background (neglecting the stochasticity here has negligible impact \cite{Figueroa:2021zah,Tomberg:2023kli}). The result \re{phi_solution} relies on the approximation that $\epsilon_2$ is constant in the constant-roll phase that follows USR. For our models, $\epsilon_2$ changes by a few percent during the constant-roll phase: it stays within 1\% of the central value for 4.1, 4.0, and 5.2 e-folds in the asteroid, solar, and supermassive cases, respectively, measured on the classical, non-stochastic trajectory. Percent level variations in $\epsilon_2$ are expected to cause percent level variations in $\zeta$.

In the $\Delta N$ formalism, the curvature perturbation $\zeta$ is equal to $\Delta N=N-\bar N$, where $N$ is the local number of e-folds and $\bar N$ is its average over a large volume, which, assuming ergodicity, we equate with the ensemble average. For the models considered here, $\bar N$ is very close to the non-stochastic classical value of $N$. (In the case of Planck-scale relics considered in \cite{Figueroa:2021zah}, this is not true.) In practice, when computing $N$, we turn off the stochastic noise at an intermediary time $\tilde N$ corresponding to a final coarse-graining scale $k=k_\sigma(\tilde N)$ and evolve classically until the end-of-inflation surface. Since our local background quantities are coarse-grained with a top-hat window function in $k$-space, we interpret the $\Delta N$ values that we obtain to be similarly coarse-grained, and write
\bea \label{zetacoarse}
  \Delta N_{<k}(\bx) = \zeta_{<k}(\bx) \equiv \int \frac{\rmd^3 p}{(2\pi)^{3/2}} \zeta_\bp e^{i\bp\cdot\bx} \theta(k-p) \ ,
\eea
where $\zeta_\bp$ are the individual Fourier modes and $p\equiv|\bp|$. The position vector $\bx$ labels the different coarse-grained patches, corresponding to the patch's centre, similarly to the coarse-grained field $\bar\phi$. Without loss of generality, we set $\bx=0$.\footnote{We assume that the different patches are independent of each other and don't overlap, that is, the patches are much smaller than their separation. This is a good approximation when PBHs are rare.}

In the constant-roll approximation with the trajectories \re{phi_solution}, we obtain $\Delta N_{<k}=\zeta_{<k}$ in terms of $X_{<k}$ as \cite{Tomberg:2023kli}
\bea \label{zeta_vs_X}
    \zeta_{<k} = - \frac{2}{\epsilon_2} \ln( 1 - \frac{\epsilon_2}{2} X_{<k} ) \ .
\eea
Varying $k$ but keeping the noise realisations $\hxi_k$ in $X_{<k}$ fixed, we obtain $\zeta$ coarse-grained over different momentum shells in the same patch. Assuming spherical symmetry around $\bx=0$, we have $\zeta_\bp$ = $\zeta_p$, so \re{zetacoarse} gives 
\bea \label{zeta_k}
  \zeta_k = \sqrt{\frac{\pi}{2}} \frac{1}{k^3} \frac{\rmd \zeta_{<k}}{\rmd\ln k} \ .
\eea
With $\zeta_k$ we can build the radial profile in position space using inverse Fourier transform. The problem of solving the stochastic equations and finding the profile has then been reduced to calculating sums of Gaussian random variables in $X_{<k}$. This can reduce the computation time by an order of magnitude or more over a brute force solution of the stochastic differential equations, depending on the case. The details of the inflaton potential are encapsulated in the power spectrum $\Pzeta(k)$ and the slow-roll parameter $\epsilon_2$, which controls the level of non-Gaussianity. 

We note that using the $k$-space top-hat window function causes some difficulty in interpreting the above steps. In particular, it is not clear at which point we should move from momentum space to position space. We do this at the end, calculating $\zeta$ in momentum space and then doing the inverse Fourier transform. However, we could also choose to take already $\bar\phi$ or $X_{<k}$ to be in position space, and then compute $\Delta N$ and $\zeta$ through the position space counterpart of the relation \re{zeta_vs_X}. The root of the ambiguity is the fact that we model a localised patch of space with a quantity that is coarse-grained in momentum space. The issue could be resolved by instead considering a window function localised in real space, but as mentioned above, this would make the stochastic analysis considerably more difficult.

\subsection{Compaction function} \label{sec:com}

Knowing the coarse-grained $\zeta$ within a patch is not enough to determine whether it will collapse into a PBH. Relativistic spherically symmetric simulations with smooth radial profiles have shown that the maximum value of $\zeta$ does not strongly correlate with PBH formation, but a good indicator of collapse is provided by the maximum value of the compaction function \cite{Shibata:1999zs, Harada:2013epa, Harada:2015yda, Musco:2018rwt, Young:2019yug, Young:2019osy, Escriva:2019nsa, Escriva:2019phb, Atal:2019erb, Yoo:2020lmg, Escriva:2020tak, Musco:2020jjb, Escriva:2021aeh, Escriva:2022pnz, Yoo:2022mzl, Harada:2023ffo, Escriva:2023uko, Escriva:2023qnq}
\bea \label{Cdef}
  \C(t, r) \equiv \frac{2\GN\Delta M(t, r)}{R(t, r)} \ ,
\eea
where $\GN$ is Newton's constant\footnote{In our units, $2\GN=1/(4\pi)$, but we prefer to keep $\GN$ explicit here to avoid confusion.}, $\Delta M(t,r)$ is mass excess over the background, and $R(t,r)$ is the areal radius at time $t$ and coordinate radius $r$. In the gradient approximation, on super-Hubble scales, the metric within one patch can be written as
\bea
  \rmd s^2 = - \rmd t^2 + a(t)^2 e^{2\zeta(r)} ( \rmd r^2 + r^2 \rmd\Omega^2 ) \ ,
\eea
so $R(t, r)=a(t) r e^{\zeta(r)}$. In the same approximation, $\C$ is independent of time and given in terms of $\zeta$ as
\bea \label{C}
  \mathcal C(r) &=& \frac{2}{3} [ 1 - ( 1 + r \zeta' )^2 ] \ ,
\eea
where prime denotes derivative with respect to $r$. The prefactor is gauge-dependent \cite{Shibata:1999zs, Harada:2015yda, Yoo:2022mzl, Harada:2023ffo}. It has been argued that the average of the compaction function is an even better indicator of collapse \cite{Escriva:2019phb, Atal:2019erb, Escriva:2020tak, Escriva:2021aeh, Escriva:2022pnz, Escriva:2023uko}. As in \paperI, we define the average compaction function as
\bea \label{avC}
  \avC(r) &\equiv& \frac{3}{R^3} \int_0^{R} \rmd \tilde R \tilde R^2 \C = - \frac{2}{r^3 e^{3\zeta(r)}} \int_0^r \rmd \tilde r \tilde r^2 e^{3\zeta} [ 2 \tilde r \zeta' + 3 ( \tilde r \zeta' )^2 + ( \tilde r \zeta' )^3 ] \ ,
\eea
where in the second equality we have used the gradient approximation \re{C}. The definition of $\avC$ assumes that $R$ is monotonic, \ie that perturbations are of type I \cite{Kopp:2010sh, Harada:2013epa, Carr:2014pga, Uehara:2024yyp, Harada:2024jxl, Shimada:2024eec, Escriva:2025rja, Uehara:2025idq}. Note that $\avC(r)$ is not the average over the proper volume, but over the volume defined by the areal radius $R$. In other studies, the integral has been extended to the maximum of $\C(r)$ (in the cases considered $\avC$ before \paperI, $\C(r)$ had only one maximum). When stochastic effects are taken into account, $\C(r)$ has a large number of maxima. We therefore leave the integration endpoint free, and find the global maximum of the function $\avC(r)$. We denote the location of the global maximum of $\C(r)$ and $\avC(r)$ by $\rmax$ (it should be clear from the context which function is meant).

The definition of the compaction function in terms of mass excess requires specifying the background mass. See \cite{Harada:2023ffo} for different definitions of the compaction function, corresponding to different spatial slicings and background masses. The standard definition (which gives the above result) involves only the first derivative of $\zeta(r)$, whereas some other definitions lead to dependence on the second derivative. Correspondingly, the average compaction function would involve the first derivatives, like our $\C$. For the smooth profiles used in the literature, this does not make much difference, but for our stochastic profiles, differentiation enhances the spikiness, and could lead to a much higher abundance. The collapse thresholds in simulations have been calculated using the standard definition of the compaction function above.

To calculate the compaction function, we need the radial profile $\zeta(r)$. Works before \cite{Tada:2021zzj} (the first paper to calculate the compaction function from stochastic inflation) and \paperI\ either used ad hoc profiles \cite{Shibata:1999zs, Harada:2013epa, Nakama:2013ica, Harada:2015yda, Musco:2018rwt, Kawasaki:2019mbl, Young:2019yug, Kalaja:2019uju, Escriva:2019nsa, Escriva:2019phb, Yoo:2020lmg, Escriva:2022pnz, Germani:2023ojx, Escriva:2023qnq} or mean profiles derived assuming Gaussianity (sometimes with a non-Gaussian component) \cite{Germani:2018jgr, Young:2019osy, Atal:2019cdz, Kalaja:2019uju, Atal:2019erb, Musco:2020jjb, Kitajima:2021fpq, Yoo:2022mzl, Escriva:2023uko}. We showed in \paperI\ that the profiles we calculate from first principles taking into account stochastic effects are very different from these profiles. Following the discussion in the last section, we calculate the profile $\zeta(r)$ by inverse Fourier transform
\bea \label{zeta}
  \!\!\!\!\!\!\!\!\!\! \zeta(r) &=& \frac{1}{(2\pi)^{3/2}} \int \rmd^3 k \, \zeta_\bk e^{i\bk\cdot\bx} = \sqrt{\frac{2}{\pi}} \int_0^{\infty} \rmd k \, k^2 \zeta_k \frac{\sin(k r)}{k r} = \int_0^{\infty} \frac{\rmd k}{k} \, \frac{\rmd \zeta_{<k}}{\rmd \ln k} \frac{\sin(k r)}{k r} \ ,
\eea
where we have assumed spherical symmetry and used \re{zeta_k}. Note that $\bx$ plays a different role in \re{zeta} than in \re{zetacoarse}. In \re{zetacoarse}, it labeled the different patches, so its value was constant (and chosen to be zero) when we computed the Fourier modes within one patch. In \re{zeta}, $\bx$ denotes position within one patch. Discretizing $k$ and recalling that $\zeta$ is a stochastic function, we now use \re{X_solution}, \re{zeta_vs_X}, and It\^{o}'s lemma \cite{Ito:1944, Vennin:2015hra, Tomberg:2023kli} for the differentiation of stochastic functions to obtain\footnote{\paperI\ has a sign error, with the opposite sign for the second term in \re{zeta_2} and \re{r_zeta_prime}. To apply the corresponding equation in \paperI\ correctly, we should drop the last $k$ mode in the sum \re{X_solution}, so that the same $\hxi_k$ does not feature both in the numerator and denominator in the first term in \re{zeta_2}, \re{r_zeta_prime}. This is how stochastic variables are usually handled in It\^{o} calculus. If we include the last mode in the sum (as we do here), we have to flip the sign to obtain the same result. The equivalence of the two forms can be seen by expanding the first term in \re{r_zeta_prime} with respect to $\hxi_k$ and replacing $\hxi_k^2$ by its expectation value of unity. The results presented in this paper are based on the same algorithm as in \paperI, which includes the wrong sign. However, the quantitative impact is small, relevant mostly at small $r$ values, and the effect on the maxima of the compaction function is negligible. For a detailed discussion of It\^{o} calculus in stochastic inflation, see \cite{Tomberg:2024evi}.}
\begin{equation} \label{zeta_2}
    \zeta(r) = \sum_{k=\kini}^{\kend} \Bigg[ -\frac{\hxi_k}{1-\frac{\epsilon_2}{2}X_{<k}}\sqrt{\Pzeta(k) \, \rmd \ln k}  - \frac{\epsilon_2}{4\qty(1-\frac{\epsilon_2}{2}X_{<k})^2} \Pzeta(k) \, \rmd \ln k \Bigg] \frac{\sin{}(k r)}{k r} \ ,
\end{equation}
where $\kend$ is the largest wavenumber considered. This can correspond to the last mode that exits the coarse-graining scale $\kc$ during inflation, but the results are not sensitive to the precise choice of $\kend$, as long as $\kend\gg\kpeak$, so that the amplitude of the neglected modes is small (see appendix \ref{app:res} for details). Similarly, the results are not sensitive to $\kini$, as discussed above. In our simulations, we consider $4 \times 10^4$ $k$-modes per patch for each mass case.

Differentiating \re{zeta_2}, we get the function $r\zeta'$ that appears in $\C$:\footnote{In \cite{Tada:2021zzj, Animali:2025pyf}, the authors built $r\zeta'(r)$ using a ``coarse-shelled curvature perturbation''. The approach is similar to ours: the authors also compute $\zeta$ coarse-grained over different scales in the same patch and deduce the profile from there. However, they equate $r$ directly with wavenumber as $k\sim1/r$ and use an approximation to estimate $r\zeta'(r)$. We perform the full inverse Fourier transform, including higher momenta where the Fourier transform kernel function $\cos(kr)-\frac{\sin(kr)}{kr}$ also has support.}
\bea \label{r_zeta_prime}
    r\zeta'(r) &=& \sum_{k=\kini}^{\kend} \Bigg[ -\frac{\hxi_k}{1-\frac{\epsilon_2}{2}X_{<k}}\sqrt{\Pzeta(k) \, \rmd \ln k}  - \frac{\epsilon_2}{4\qty(1-\frac{\epsilon_2}{2}X_{<k})^2} \Pzeta(k) \, \rmd \ln k \Bigg] \el
    && \times\left[\cos{} ( k r ) - \frac{\sin{}(k r)}{k r}\right] \ .
\eea
In practice, for better numerical accuracy, we use an improved version of this sum with better convergence properties, discussed in appendix \ref{app:res}. We then straightforwardly obtain $\C(r)$ and $\avC(r)$ from \re{C} and \re{avC}, separately for each patch.

\subsection{PBH mass}

We assume that a PBH forms if the maximum of $\C(r)$ or $\avC(r)$ exceeds the collapse threshold denoted by $\Cth$ or $\avCth$ (discussed below in \sec{sec:sim}), and all mass inside $\rmax$ ends up in the PBH when the corresponding scale re-enters the Hubble radius. The PBH mass at formation is then, taking the mass excess from \re{Cdef} (see \eg \cite{Carr:1975qj, Tomberg:2024chk}),
\bea \label{mass}
  M_{\rmax} &=& ( 1 + \Cmax ) \bar M_{\rmax} = 5.6 \times 10^{15} ( 1 + \Cmax ) (k_* \rmax)^2 M_\odot \ ,
\eea
where $\bar M_{\rmax}$ is the mass contained in the FLRW region with coordinate radius $\rmax$, $\Cmax$ is the global maximum value of $\C(r)$, and $k_*=0.05$ Mpc$^{-1}$ is the CMB pivot scale. We assume Standard Model particle content and standard thermal history. Note that we match the mass at background coordinate radius, not physical radius. Relatedly, there is a difference between the Misner--Sharp mass and the energy density integrated over the volume when gradients are taken into account. However, $\zeta$ is at most of order unity, so the difference between the physical radius $R(t,r)/a(t) = r e^{\zeta(r)}$ and the coordinate radius $r$, and the Misner--Sharp mass and the integrated energy density, is not large. We do not include the conventional multiplicative factor $\c\approx0.2$ \cite{Carr:1975qj} that characterises the efficiency of the collapse. The stochastic kicks have a large effect on the mass via variation of $\rmax$, and their effect on the collapse dynamics introduces further uncertainty much larger than order one corrections.

We assume that all mass inside the highest maximum of $\C(r)$ collapses into a PBH and all matter further out disperses, even if $\C(r)$ exceeds the collapse threshold also there. Instead, it could happen that a PBH forms later on those larger scales. (It has also been argued in \cite{Atal:2019erb} that a PBH only forms at the smallest radius where $\C(r)>\Cth$; see \cite{Nakama:2014fra, Escriva:2023qnq} for findings to the contrary and discussion of the dependence on the profile). Our assumption is conservative in the sense that it provides a lower limit to the PBH mass. Also, as our resolution is worse at large $r$, the identification of $\Cmax$ values is less robust there. As we have the full radial profile for each patch, the peak-in-a-peak problem is solved, once we know which prescription to use for nested peaks.

We also consider the effect of critical behaviour. For masses near the threshold, the compaction function can be almost constant for some time as the system shrinks while shedding mass, and the final mass is given by the critical scaling law \cite{Niemeyer:1997mt, Gundlach:1997nb, Yokoyama:1998xd, Niemeyer:1999ak, Gundlach:2002sx, Musco:2004ak, Musco:2008hv, Musco:2012au, Baumgarte:2015aza, Kuhnel:2015vtw, Germani:2023ojx}
\bea \label{criticalmass}
  M_{\text{crit}} = M_{\rmax} K ( \Cmax - \Cth )^{\c} \ ,
\eea
where $M_{\rmax}$ is given in \re{mass}, and $K$ and $\c$ are constants. The critical exponent $\c$ has been found to be rather independent of the profile, $\c=0.36$ \cite{Evans:1994pj, Koike:1995jm, Maison:1995cc, Gundlach:1997nb, Gundlach:1999cw, Niemeyer:1999ak, Gundlach:2002sx, Musco:2008hv, Musco:2012au, Escriva:2019nsa, Germani:2023ojx, Ianniccari:2024ltb} (see \cite{Escriva:2023uko} for a study involving non-Gaussianity). The prefactor depends on the profile: we adopt $K=4$, found in \cite{Escriva:2021aeh} in the limit of high peaks for $\Cth=0.4$, although its applicability to our case is unclear. It is also not clear how far the critical scaling extends. The scaling is expected to hold only in the vicinity of the threshold, and deviations start to appear at $\Cmax-\Cth\approx10^{-2}$ \cite{Musco:2008hv}, although at least for some profiles the deviations are only of the order 15\% even for large values \cite{Escriva:2019nsa}. We consider two alternatives. One is to take a narrow range of applicability, where we cut the scaling off at $\Cmax-\Cth=10^{-2}$ (above this, we use \re{mass} for the mass), the other is to extend the scaling to all values of $\Cmax$. The extended range is unlikely to be realistic, but it gives an upper estimate of the impact of critical scaling.

\subsection{Window function and transfer function}

A window function has usually been inserted (and sometimes presented as necessary) in the expression \re{zeta} for $\zeta$ to cut off the integral at large $k$ values and define the PBH mass by introducing a scale \cite{Ando:2018qdb, Young:2019osy, Kalaja:2019uju, Franciolini:2023wun, Pi:2024ert}. However, its physical motivation is unclear. In \cite{Kalaja:2019uju} it was argued that a window function is required to get a differentiable field. But inflation generates a white noise random field in momentum space, not position space: the inverse Fourier transform of this field is differentiable, as non-differentiability in momentum space is not mapped onto non-differentiability in position space but instead onto behaviour at large separations (following the convergence properties of the inverse Fourier transform). In our case the effects of small-scale modes are suppressed by the fact that the power spectrum has a well-defined peak and falls off rapidly for large wavenumbers.

We do not introduce any window function when calculating $\zeta$.\footnote{The window function discussed here should not be confused with the window function introduced in \re{dec} to separate the small and long wavelength inflaton modes, nor with the integration kernels used to calculate position space functions like $\zeta(r)$ in \re{zeta_2} and $r\zeta'(r)$ in \re{r_zeta_prime}. The former is an inevitable part of the definition of perturbation theory, which involves a division between the background and perturbations, although the precise split (\ie the choice of window function) is dictated by convenience. The latter are part of the definition of the inverse Fourier transform.} The PBH mass is given in \re{mass} by the location $\rmax$, whose distribution is determined by the stochastic process, not given a priori. We have checked that in the asteroid case a Gaussian window function would reduce the PBH abundance by a factor of 2 for $\Cth=0.4$ and a factor of 6 for $\avCth=0.4$.

There is also a physical smoothing effect due to sub-Hubble evolution, described by the transfer function. We calculate $\zeta(r)$ in the regime where all the relevant scales are very super-Hubble. Modes start evolving when they enter the Hubble radius, and in the linear regime their amplitude is suppressed by the radiation era transfer function
\bea \label{T}
  T(k) = 3 \frac{ \sin x - x \cos x }{x^3} \ ,
\eea
where $x\equiv k/(\sqrt{3} a H)$. Our stochastic profiles involve a wide range of scales, and much of the stochastic structure on small scales will be smoothed before the scale $\kpeak$ starts evolving. When using the transfer function, we fix $aH=\kpeak$, assuming that PBH collapse is determined by the conditions when the scale $\kpeak$ enters the Hubble radius, and that the evolution until then is captured by the linear radiation era transfer function. This is a rather crude approximation. Given the way we determine the PBH mass, it might make more sense to use $1/\rmax$ as the smoothing scale, but this quantity is only determined after the smoothing. The peak of the mass function corresponds to scales larger than $1/\kpeak$. To go beyond the linear smoothing \re{T}, we would need to consider the complicated non-linear evolution after Hubble entry. For discussion of the non-linear transfer function, see \cite{Kalaja:2019uju, DeLuca:2023tun, Franciolini:2023wun}. A reliable calculation of the abundance will require relativistic hydrodynamical collapse simulations with stochastic profiles, although even then resolving the evolution of the various modes seems difficult given the range of scales involved. In the present paper, we simply use the linear transfer function. We compare the case without the transfer function and the case where we replace $\zeta_k\to T(k)\zeta_k$ when calculating $\zeta(r)$ from \re{zeta}. We identify quantities computed with the transfer function with the subscript $T$, for example $\zeta_T$ and $\C_T$.

\subsection{Simulation procedure} \label{sec:sim}

Our simulations proceed as follows. (The algorithm is spelled out in appendix C of \paperI.) Having chosen the inflationary potential and calculated the power spectrum $\Pzeta(k)$ and the value of $\epsilon_2$, we decide on the values $\kini$ and $\kend$ and draw the amount of random numbers $\hxi_k$ determined by the $k$ resolution we consider. For the simulations considered in this paper, $\kini = k_*=0.05$ Mpc$^{-1}$, and $\kend$ is a mode that crosses the Hubble radius 10 e-folds before the end of inflation. The mode resolution is $\rmd \ln k = 10^{-3}$, which translates to about $4\times10^4$ modes per simulation. We then calculate $\zeta(r)$, $\C(r)$, and $\avC(r)$ as discussed above, separately with and without the transfer function. We record their maximum values, denoted by $\zetamax$, $\Cmax$, and $\avCmax$, respectively. We also record the values $\rmax$ at which $\Cmax$ and $\avCmax$ are reached, the number of peaks for which $\C(r)>\Cth$ or $\avC(r)>\avC_\thr$, and the full width at half maximum for the highest peak, $w\equiv r_2-r_1$, where $\C(r_1)=\C(r_2)=\ha \Cmax$. We also check and record whether we have $r\zeta'<-1$ anywhere, corresponding to type II perturbations \cite{Kopp:2010sh, Harada:2013epa, Carr:2014pga, Uehara:2024yyp, Harada:2024jxl, Shimada:2024eec, Escriva:2025rja, Uehara:2025idq}. To characterise the importance of departures from smoothness, we fit the function $A (1 - \tanh[B\ln(r/r_0)])$ to each realisation of $\zeta(r)$, where $A$, $B$, and $r_0$ are fitting constants. The fitting constants are set to $A = \zetamax/2$, $B = -r\zeta'/A$, and $r_0 = r e^{-\rmd\ln r}/\kpeak$, where $r$ is fixed by where $\zeta(r)$ has its steepest descent and $\dd \ln r$ is the $r$ resolution as discussed in appendix \ref{app:res}. We calculate $\Cmax$, $\avCmax$, and the associated $\rmax$ values also from this fit. We then run the simulation again, \ie we draw a new set of random variables $\hxi_k$, and repeat $10^8$ times to collect statistics.
 
In order to calculate the abundance of PBHs from the distribution of $\Cmax$ and $\avCmax$, we need to know the thresholds $\Cth$ and $\avC_\thr$ above which a PBH forms. (See \eg \cite{Helou:2016xyu} for discussion of PBH formation and trapped surfaces.) The threshold has been calculated in relativistic hydrodynamical simulations with smooth profiles as well as analytically, mostly in cases where $\C(r)$ has only one peak; more complicated profiles have been studied in \cite{Nakama:2014fra, Atal:2019erb, Escriva:2023qnq}. The threshold value has been found to be $\Cth=0.4\ldots2/3$, depending on the radial profile, or $\avC_\thr=0.4$, rather independently of the profile  \cite{Shibata:1999zs, Harada:2013epa, Nakama:2013ica, Nakama:2014fra, Harada:2015yda, Musco:2018rwt, Germani:2018jgr, Kawasaki:2019mbl, Young:2019osy, Young:2019yug, Kalaja:2019uju, Atal:2019cdz, Atal:2019erb, Escriva:2019nsa, Escriva:2019phb, Yoo:2020lmg, Escriva:2020tak, Musco:2020jjb, Escriva:2021aeh, Kitajima:2021fpq, Escriva:2022pnz, Yoo:2022mzl, Germani:2023ojx, Escriva:2023qnq, Escriva:2023uko, Kehagias:2024kgk, Ianniccari:2024bkh, Ianniccari:2024ltb}. New collapse simulations are needed to determine the collapse criterion for the profiles with the sharp fluctuations generated by the stochastic process. As high-frequency modulation of a peak is known to facilitate collapse \cite{Nakama:2014fra} and two nearby smooth peaks under the threshold individually can lead to PBH formation \cite{Escriva:2023qnq}, it is likely that the values $\Cmax$ or $\avCmax$ alone will not determine whether a PBH forms, but factors such as peak width and the distance of the peaks from each other will also play a role. Also, many of the stochastic peaks are very narrow and will hence lead to large pressure gradients once they enter the Hubble radius, which resist collapse and can smooth out the peaks before they have time to collapse. Determining the collapse criterion accurately will thus require redoing simulations of PBH formation with stochastic profiles. In the present paper, we simply find the distributions of $\Cmax$ and $\avCmax$ and compare to the case where the stochastic effects are not accounted for. When a value for the collapse threshold is needed, we adopt $\Cth=0.4$ or $\avC_\thr=0.4$, unless otherwise noted.

As points of comparison, we use results based on mean profiles. In \cite{Tomberg:2023kli}, it was shown that in the stochastic constant-roll process, for a fixed $X_{<\kend}\equiv X_0$, the mean $X_{<k}$ values are given by the noise realisations $\hxi_k = -X_0\sqrt{\Pzeta(k)\rmd \ln k}/\sigma_X^2$, where $\sigma_X^2 \equiv \int_{\kini}^{\kend} \rmd \ln k \, \Pzeta(k)$. By \re{zeta_2} and \re{r_zeta_prime}, this gives the profiles\footnote{These profiles are smooth, so have we dropped the It\^{o} contribution and transformed the sums into integrals.}
\bea
  \label{zeta_mean}
  \zeta(r) &=& \frac{X_0}{\sigma_X^2}\int_{\kini}^{\kend} \frac{\rmd \ln k \ \Pzeta(k)}{1-\frac{\epsilon_2 X_0}{2\sigma_X^2}\int_{\kini}^{k} \dd \ln \tk \ \Pzeta(\tk)} \frac{\sin(kr)}{kr} \ , \\
  \label{r_zeta_prime_mean}
  r\zeta'(r) &=& \frac{X_0}{\sigma_X^2}\int_{\kini}^{\kend} \frac{\rmd \ln k \ \Pzeta(k)}{1-\frac{\epsilon_2 X_0}{2\sigma_X^2}\int_{\kini}^{k} \dd \ln \tk \ \Pzeta(\tk)} \qty[\cos(kr) - \frac{\sin(kr)}{kr}] \ .
\eea
The function $\zeta$ peaks around $r=0$, while $r\zeta'$ peaks at a non-zero $r$. Using \re{zeta_mean} and \re{r_zeta_prime_mean}, we obtain the corresponding $\C(r)$ and $\avC(r)$. Their maxima $\Cmax$ and $\avCmax$ are then functions of $X_0$, which is a Gaussian random variable with $\expval{X_0} = 0$, $\expval{X_0^2} = \sigma_X^2$. This is enough for us obtain the probability distribution for $\Cmax$ and $\avCmax$. We do this separately for the non-Gaussian case above and for the Gaussian limit $\epsilon_2=0$. In the Gaussian limit, the form of the profile \re{zeta_mean} is well-known \cite{Dekel:1981}, and we already considered it in \paperI.

Let us now show the results for the abundance and mass distribution of our three different mass cases in turn.

\section{Results} \label{sec:res}

\subsection{Asteroid mass dark matter} \label{sec:ast}

\subsubsection{Parameters and radial profiles} \label{sec:ast_pars}

We first consider asteroid-mass PBHs, which can form all of the dark matter. The power spectrum resulting from the deterministic evolution (\ie neglecting stochastic noise) has $\kpeak=3.2\times10^{12}$ Mpc$^{-1}$, with $\mathcal P_\zeta(\kpeak)=7.3\times10^{-3}$, and $\epsilon_2=0.807$. (The power spectra in the asteroid, solar, and supermassive cases are shown in figures 6, 9, and 12 of \cite{Figueroa:2021zah}, and the potentials can also be found there.) In \cite{Figueroa:2020jkf, Figueroa:2021zah}, we took the PBHs to form instantly at the moment when the last scale to receive kicks in those simulations entered the Hubble radius. This scale corresponded to the end of ultra-slow-roll inflation; let us call it $\kold$ (in the asteroid case, we have $ \kold\approx5\kpeak$). To determine whether a PBH forms, we compared the coarse-grained curvature perturbation $\zeta_{<\kold}$ to the PBH formation threshold $\zetath=1$.\footnote{In the current work, we also include higher-$k$ modes, but roughly $\zeta_{<\kold} \approx \zeta_{<\kend} = \zeta(r=0)$; the last equality can be verified from \re{zeta}.} We also included the efficiency factor $\c=0.2$ for the mass that goes into the PBH. In linear perturbation theory, neglecting stochastic effects, this gave $M=7.2\times10^{-15} M_\odot$ and $\Omega_\text{PBH}=0.13$. Without the $\c$ factor, the mass is $3.6\times10^{-14} M_\odot$; for comparison, the mass within the Hubble radius when the scale $\kpeak$ enters the Hubble radius is $1.4\times10^{-12} M_\odot$.

\begin{figure}[th]
  \centering
  \includegraphics[scale=1]{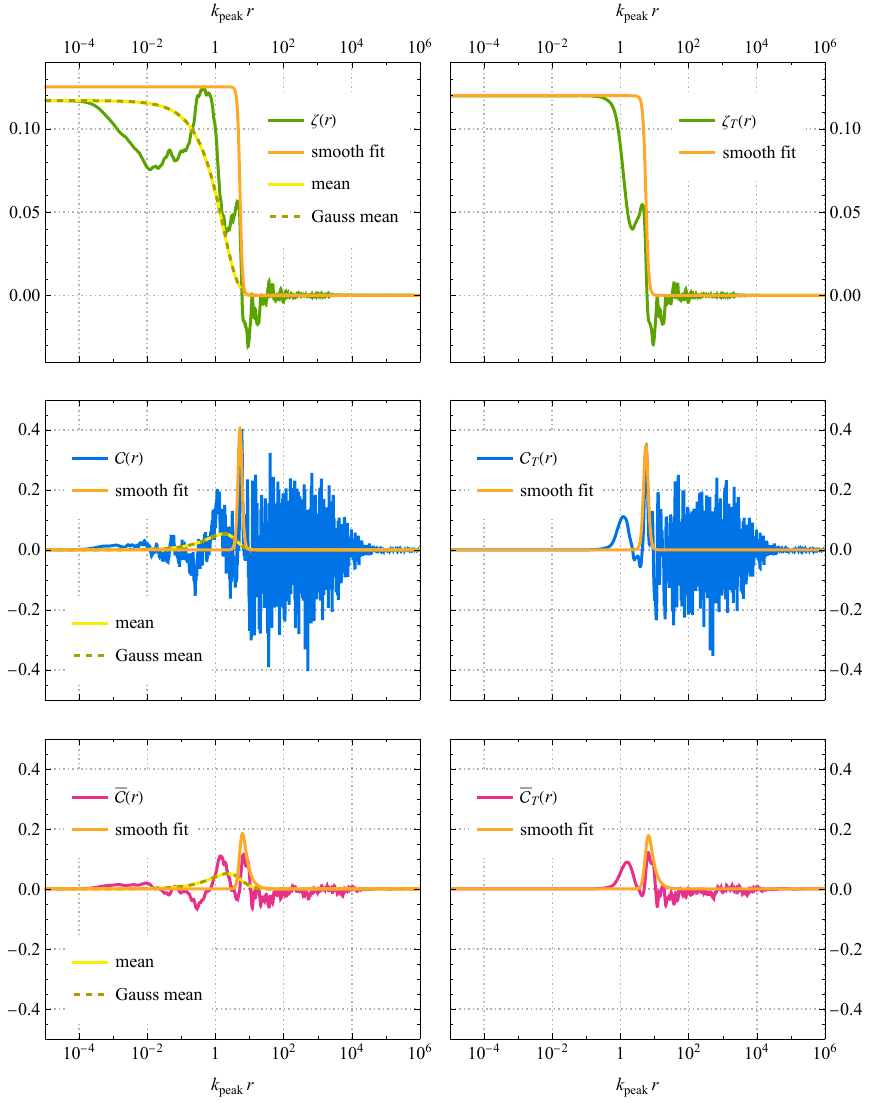}
  \caption{One realisation of the curvature perturbation $\zeta(r)$ and the corresponding compaction function $\C(r)$ and its average $\avC(r)$  in the asteroid case, together with the smooth fit and the full and Gaussian mean profiles. The right-hand figures show how small-scale variation is smoothed by the transfer function $T(k)$.}
  \label{fig:astprof}
\end{figure}

In the current setup, our $10^8$ realisations reach up to $\Cmax=0.60$ and $\avCmax=0.41$. In the top plots of \fig{fig:astprof}, we show one realisation of $\zeta(r)$, chosen so that $\Cmax=0.4$, together with the mean Gaussian and non-Gaussian radial profiles given in \re{zeta_mean}, taken to have the same value of $\zeta(0)$ as the stochastic realisation. We also show the smooth $\tanh$ fit to the profile. Below, we show the corresponding compaction function and its average in the different cases. The stochastic profiles and the fits are shown with and without the transfer function. The stochastic profiles feature many maxima, often very narrow. (See \cite{Fumagalli:2024kgg} for a different case with narrow peaks.) It was argued in \cite{Kalaja:2019uju} that having peaks inside peaks is rare because the perturbations large enough to produce PBHs are rare, but we find otherwise. As reported in \paperI, the number of peaks $p$ of $\C(r)$ above $\Cth=0.4$ in our runs follows the scaling law $n(p)\propto e^{-\a p}$ with $\a=0.8$. For discussion of the effects of resolution, see appendix \ref{app:res}. In none of the realisations do we have $r\zeta'<-1$ anywhere. Therefore the areal radius $R$ is always monotonic, and all profiles are of type I \cite{Kopp:2010sh, Harada:2013epa, Carr:2014pga, Uehara:2024yyp, Harada:2024jxl, Shimada:2024eec, Escriva:2025rja, Uehara:2025idq}\footnote{Note that whether the initial conditions are type I or II does not determine the PBH horizon structure \cite{Uehara:2024yyp, Harada:2024jxl, Shimada:2024eec, Uehara:2025idq}.}.

Comparison of the plots in \fig{fig:astprof} demonstrates how the spatial averaging smooths the compaction function, reducing the fluctuations and lowering the amplitude: the structure of $\avC(r)$ is considerably simpler than that of $\C(r)$. Taking the average has a stronger effect than including the transfer function. Without the transfer function, we have $\Cmax=0.40$ compared to $\avCmax=0.11$, and including the transfer function changes these values to $\Cmax=0.35$ and $\avCmax=0.12$. The smooth fit, in this case, identifies the highest peak in both cases, and has $\Cmax=0.40$ and $\avCmax=0.185$. With the transfer function, these values change to $\Cmax=0.35$ and $\avCmax=0.177$.

Figure \ref{fig:astprof} shows that the smooth fit reproduces the typical main features of $\zeta(r)$ (in the case when $\zetamax$ is large): central peak and decline to zero at large radii. For the stochastic profiles, the peak in $\C(r)$ corresponding to this downturn typically gives the maximum value $\Cmax$, but the fluctuations can also lead to higher peaks at larger radii. In cases where $\zeta(0)<0$ (which are not rare), the peak in $\C$ is often not related to any wide feature in $\zeta(r)$, making the usefulness of these fits limited: knowledge of the detailed peak structure is necessary. The value of $\Cmax$ determined from the fits is not very predictive of the real $\Cmax$, except at the largest values of $\Cmax$ (for which we also have the least statistics); neither is the value $\avCmax$ determined from the fit useful in predicting its real value. Taking into account the smoothing effect of the transfer function does not change these conclusions.

The mean profiles calculated from \re{zeta_mean} and \re{r_zeta_prime_mean} are even smoother than the ones obtained with the transfer function, leading to lower compaction function values, $\Cmax=0.054$ and $\avCmax=0.051$. We will see this behaviour again below when comparing the full $\C$ distributions. The Gaussian and non-Gaussian mean profiles are practically identical in this example, and the $\C$ and $\avC$ profiles are also close although they deviate for large $\C$. We haven't included the transfer function in the mean profiles, but the results with it would be similar.

It has been suggested to use the shape parameter $q\equiv-\left.\frac{r^2 \C''}{4 \C ( 1 - \frac{3}{2} \C )}\right|_{r=\rmax}$ to characterise the curvature of the $\C(r)$ profiles independently of their amplitude and to determine the PBH abundance, mostly using ad hoc radial profiles \cite{Musco:2018rwt, Escriva:2019phb, Musco:2020jjb, Escriva:2021pmf, Germani:2023ojx, Ianniccari:2024bkh}. For our stochastic profiles, the value of $q$ is practically random due to rapid variation of the profiles. This is suggested by the visual impression of \fig{fig:astprof}, and we have confirmed it by calculating the correlation between $q$ and other parameters like $\Cmax$: $q$ contains no useful information. (The value of $q$ determined from the smooth fit is less uninformative: small values of the fit $q$ correspond to larger values of $\Cmax$ and $\avCmax$, and therefore to a larger PBH abundance.) The value of $q$ is typically very large, $q\sim10^3\ldots10^4$. Including the transfer function smooths out rapid variations, so $q$ no longer peaks at $\sim10^4$, but instead its probability distribution drops monotonically, though the distribution has significant support to around $q\sim10^2$. The value of $q$ also remains uncorrelated with $\Cmax$ and $\avCmax$. For smooth profiles, the collapse threshold $\Cth$ is known to depend on $q$ and approach $2/3$ for large $q$. As the applicability of this result to our stochastic profiles is unclear, we do not vary $\Cth$ with $q$. 

\subsubsection{Abundance}

\begin{figure}[th]
  \centering
  \includegraphics[scale=1]{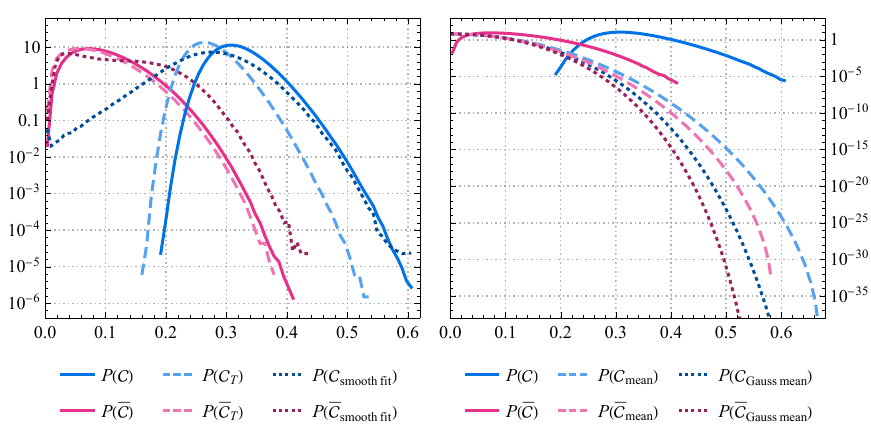}
  \caption{Left: The probability distributions of $\Cmax$ and $\avCmax$ in the asteroid case without the transfer function $T(k)$ (solid), with $T(k)$ (dashed), and from the smooth fit (without $T(k)$) (dotted), all based on the full stochastic calculation. Right: The probability distributions of $\Cmax$ and $\avCmax$ in the asteroid case from the full stochastic calculation (solid), from the mean non-Gaussian profile \re{r_zeta_prime} (dashed), and the mean Gaussian profile (dotted), all without $T(k)$.}
  \label{fig:astP}
\end{figure}

On the left in \fig{fig:astP} we show the probability distributions of $\Cmax$ and $\avCmax$ for the stochastic case, with and without the transfer function. The distributions are probably not reliable for small values of $\Cmax$, as the assumption of spherical symmetry is particularly questionable for such patches, but they anyway do not form PBHs. We also show the probability distribution we inferred from the smooth fit profiles for $\avCmax$. A key quantity is the initial fraction of patches that form PBHs\footnote{In our earlier papers, we included in the definition of $\beta$ the conventional factor of 2 used in the Press--Schechter prescription to solve the cloud-in-a-cloud problem \cite{Press:1973iz}. This factor is appropriate when using the threshold $\zetath$, because $\zeta$ takes both positive and negative values, but as $\Cmax$ is non-negative (for $0>r\zeta'>-2$), it is not needed here. Accordingly, in the current paper we have divided the values of $\b$ quoted from \paperI\ by 2.},
\begin{equation}
  \beta\equiv\int_{\Cth}^{2/3}\rmd\Cmax P(\Cmax) \ .
\end{equation}
For $\Cth=0.4$, the full stochastic profiles give $\b=0.02$. Taking $\Cth=0.5$ decreases the abundance to $\beta=1\times10^{-4}$. Using the average compaction function instead, $\avCth=0.4$, gives $\beta=1\times10^{-8}$. The prefactor is not fully resolved by our simulations, but we expect the order of magnitude to be reliable; for discussion of resolution and convergence see appendix \ref{app:res}. As our ensemble includes black holes that form at different times and the PBH fraction evolves in the radiation-dominated background, this definition of $\b$ does not directly correspond to the PBH mass fraction at any fixed time. Nevertheless, it is a convenient quantity to describe the PBH statistics. We discuss the PBH density parameter $\Omega_\text{PBH}$ in section \ref{sec:asteroid_mass_distribution}.

As we saw in the single realisation of figure \ref{fig:astprof}, including the transfer function \re{T} smooths the profile, leading to smaller values of $\Cmax$ and $\avCmax$, as shown in \fig{fig:astP}. The distribution of $\Cmax$ is shifted to significantly smaller values when $T(k)$ is included. For $\avCmax$ the effect is much more modest, as we saw in the case of the single realisation. That $\avC$ is less affected by the transfer function is unsurprising, given that the averaging already provides smoothing. For $\Cth=0.4$, the transfer function reduces the abundance $\b$ by two orders of magnitude and for $\Cth=0.5$ by three orders of magnitude. Comparing to the mean and Gaussian results (see below), we see the stochastic enhancement due to the spikes is not wiped out by the smoothing given by the transfer function. Even without the transfer function, $\avCmax=0.4$ is at the limit of our sample, so when we include the transfer function, there are no realisations with $\avCmax>0.4$. The distribution of $\avCmax$ obtained from the smooth fits has roughly the same, but slightly less steep, slope for large values as the distribution determined from the full profiles, but the normalisation is off by an order of magnitude. For $\Cmax$ the distribution determined from the smooth fits agrees reasonably well with the full distribution for large values (the flattening at the end is likely a numerical artifact).

On the right in \fig{fig:astP}, we compare the full $\Cmax$ and $\avCmax$ distributions to those calculated from the mean profile \re{zeta_mean} in the non-Gaussian case (with $\epsilon_2\neq0$) and in the Gaussian case (with $\epsilon_2=0$), all without the transfer function. We noted in \paperI\ that the Gaussian mean curve does not reproduce the probability distribution anywhere, and the same is true for the non-Gaussian mean. The Gaussian distributions fall more steeply than the non-Gaussian ones, showing how the non-Gaussianity enhances the abundance even for the mean profile. We discuss this in more detail in \sec{sec:disc}. The higher the collapse threshold, the larger the enhancement over the mean profile result, so in this sense our assumption $\Cth=0.4$ is conservative. However, we expect the collapse threshold to be higher in the stochastic case than for the smooth mean profile, as larger pressure gradients resist collapse, so using the same threshold for both may be misleading.

In the Gaussian case, using the mean profile gives $\b=4\times10^{-15}$ for $\Cth=0.4$, $\b=2\times10^{-26}$ for $\Cth=0.5$, and $\b=8\times10^{-18}$ for $\avCth=0.4$. Comparing to the earlier numbers, the Gaussian mean profile underestimates our stochastic result by 13 to 22 orders of magnitude. These numbers can also be compared to the Gaussian case calculation with the threshold $\zetath=1$, which gives $\b=1\times10^{-16}$; including the stochastic tail as in \cite{Figueroa:2020jkf, Figueroa:2021zah} raises this by 5 orders of magnitude to $\b=8\times10^{-12}$. The full stochastic enhancement compared to the Gaussian $\zetath=1$ case is 8 to 14 orders of magnitude, depending on the collapse threshold. More abundance numbers are collected in table \ref{tab:numbers} in terms of $\Omega_\text{PBH}$, which we will discuss in the next section.

In the above case, taking the collapse threshold to be set by the compaction function instead of the curvature perturbation $\zeta$ increases the abundance when stochastic fluctuations in the profile are not accounted for, but this does not always happen. In \cite{Kawasaki:2019mbl} it was found that using $\Cmax$ as the collapse criterion (with the Gaussian mean profile) instead of $\zetamax$ suppressed the PBH abundance by a factor of a few, growing to about $10^8$ when including skewness. In \cite{Young:2019yug} suppression by many orders of magnitude was also found.

\begin{figure}[th]
  \centering
  \includegraphics[scale=1]{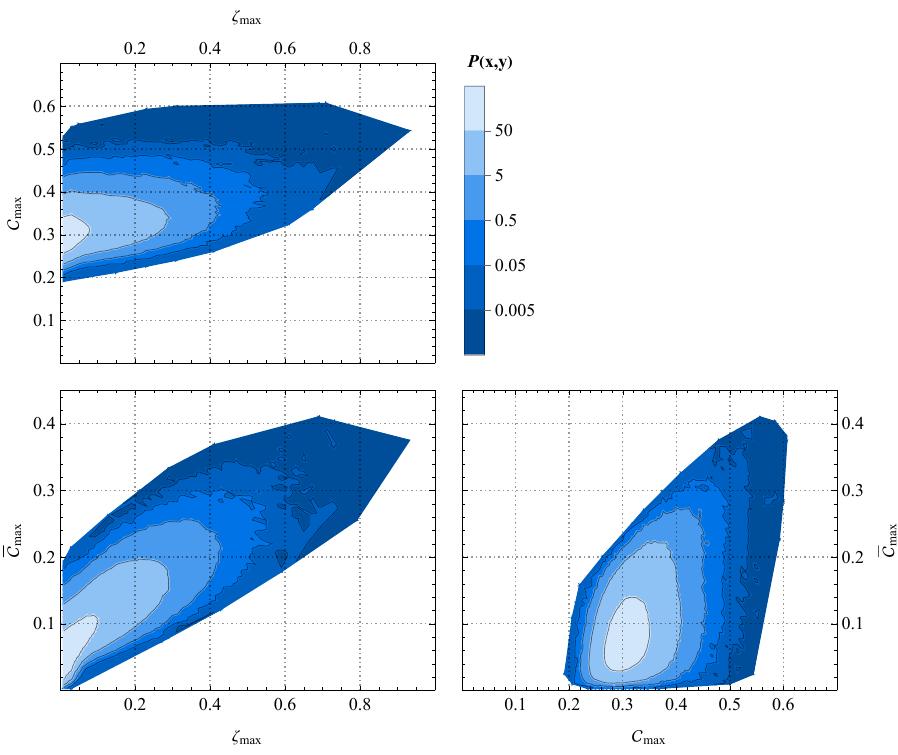}
  \caption{Probability distributions of $\Cmax$, $\avCmax$, and $\zetamax$ in the asteroid case.}
  \label{fig:astcor}
\end{figure}

In \fig{fig:astcor} we show the correlations between $\Cmax$, $\avCmax$, and $\zetamax$. The last is a proxy for $\Delta N$ from other stochastic studies. Large values of $\avCmax$ correspond to large values of $\Cmax$, but not vice versa; over a large range of values, $\Cmax$ and $\avCmax$ are almost uncorrelated. Large values of $\avC$ typically come from a profile where $\C$ is large over a wide range of $r$ values, instead of having a single narrow peak. However, both the smallest and the largest values of $\Cmax$ and $\avCmax$ correspond to moderate, not large, values of the width $w$ (for $\C$, the width peaks around $\Cmax=0.27$). As noted in \paperI, $\Cmax$ and $\zetamax$ are poorly correlated, and the value of $\Cmax$ cannot be determined from the value of $\zetamax$. This could call into question the applicability of the relation between the collapse criteria for $\C$ and $\zeta$ found with coarse-grained shells in \cite{Tada:2021zzj, Animali:2025pyf}. However, $\avCmax$ and $\zetamax$ do show clear correlation. It remains to be checked with collapse simulations whether $\Cmax$ or $\avCmax$ or neither is a good indicator of PBH formation for very spiky profiles.

\begin{figure}[th]
  \centering
  \includegraphics[scale=1]{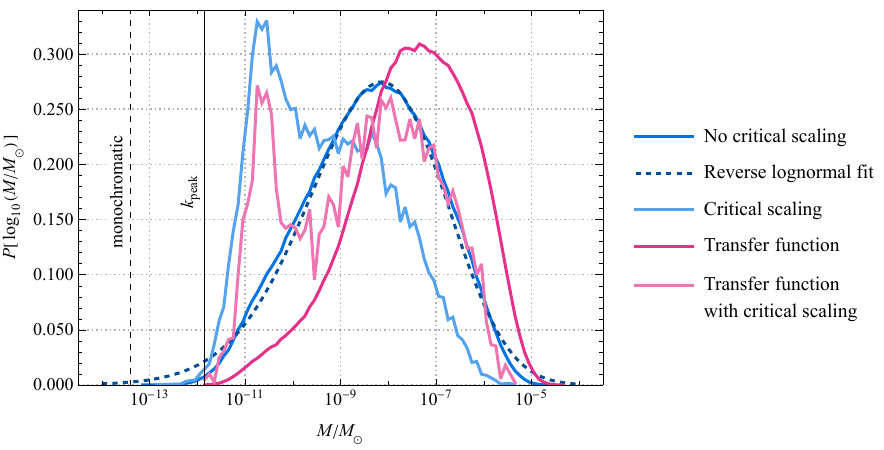}
  \caption{The mass distribution in the asteroid case with the collapse threshold $\Cth=0.4$ for five cases: without critical scaling, the reverse lognormal distribution fit (without critical scaling), with critical scaling (for $0.4 < \Cmax < 0.41$), with the transfer function but no critical scaling, and with the transfer function and critical scaling. The vertical lines mark the monochromatic mass considered in \cite{Figueroa:2020jkf, Figueroa:2021zah} and the mass corresponding to $\kpeak$, at $3.6\times10^{-14}M_\odot$ and $1.4\times10^{-12}M_\odot$, respectively.}
  \label{fig:astmass}
\end{figure}

\subsubsection{Mass distribution} \label{sec:asteroid_mass_distribution}

\noindent In \fig{fig:astmass} we show the mass distribution of the stochastic computation, calculated assuming $\Cth=0.4$, both with and without the effect of critical collapse \re{criticalmass} and with and without the transfer function \re{T}. Calculating the mass distribution involves a finer splitting than finding the number density, and not all mass bins are well sampled. Furthermore, the distribution extends to higher and higher masses with increasing resolution, as more peaks at large $\rmax$ are resolved (we discuss resolution issues in appendix \ref{app:res}). This is also why the peak mass is higher here than in the plot of \paperI: the mode resolution of the present simulations is four times higher. Unlike for the abundance (for which the relative error decreases with resolution and the order of magnitude is well determined) the mass does not show signs of convergence. This is also true in the solar and the supermassive case. The growth of computational time due to increasing the resolution is prohibitive even with our approximation that simplifies the solution of the stochastic equations to an algebraic combination of Gaussian random variables. This is also true in the solar and supermassive cases. So while we can robustly say that the stochastic effects increase the peak value and the width of the mass distribution, our quantitative results for this growth are only indicative.

In the mass distribution shown in \fig{fig:astmass}, the 95\% range $9 \times 10^{-12}\ldots1 \times 10^{-6} M_\odot$ covers 5 orders of magnitude, mostly above the asteroid range. (Microlensing constraints put the upper end of the allowed mass window at $10^{-11} M_\odot$, one order of magnitude below the mass of Vesta, the most massive asteroid.) The potential was originally tuned (neglecting stochastic effects) to produce PBHs with masses close to the lower end of the asteroid mass range, $3.6\times10^{-14} M_\odot$, but practically all of the masses are larger than this. The masses are also larger than the Hubble mass $1.4\times10^{-12} M_\odot$ corresponding to the scale $\kpeak$, as also found in \cite{Gow:2020bzo}. This mass function is much wider than the lognormal mass distribution and other extended mass functions considered in the literature, including the stochastic calculation of \cite{Tada:2021zzj}, and the observational constraints should be reworked \cite{Green:2016xgy, Carr:2017jsz, Gow:2020bzo, Gow:2020cou, Gorton:2024cdm}. It is approximately fit by the reverse lognormal distribution $e^{-[\ln(c-x)-\mu]^2/(2\sigma^2)}/[(c-x)\sigma\sqrt{2\pi}]$, but for $x = \log_{10}(M/M_\odot)$ instead of $M/M_\odot$, with $\mu = 2.569$, $\sigma = 0.112$, and $c = 4.746$, although the fit overestimates both the high- and the low-mass tail.

Including critical collapse shifts the distribution to smaller masses, as shown in \fig{fig:astmass} (the fine structure seen in the plot is due to sampling issues, the physical curve is smooth). The 95\% mass range becomes $5\times10^{-12}\ldots2\times10^{-7} M_\odot$ for the narrow width of applicability of the critical scaling. The case where the critical scaling applies to all masses is virtually identical. Using the transfer function shifts the distribution to larger masses, because it smooths out small-scale perturbations. The 95\% mass range (without critical scaling) shifts upwards by less than an order of magnitude, to $6\times10^{-11}\ldots4\times10^{-6} M_\odot$, but still covers five orders of magnitude. When both critical scaling and the transfer function are included, their effects to some extent cancel each other, and the 95\% mass range is close to the case when neither is included.

If all PBHs had the same mass $M$, their contribution to the energy density today would be (for $H_0=70$ km/s/Mpc) \cite{Tomberg:2024chk}
\bea \label{omega_M}
  \Omega_{\text{PBH}}(M) = 9.16 \times 10^7 \b \left(\frac{M}{M_\odot}\right)^{-1/2} \ .
\eea
For the PBH mass distribution $P[\log_{10} (M/M_\odot)]$ (normalised to unity, $\int_0^\infty \rmd x P(x)=1$), the total PBH density parameter is given by the integral
\bea \label{omega_total}
  \Omega_{\text{PBH}} = 9.16 \times 10^7 \int_0^\infty \rmd [\log_{10} (M/M_\odot)] \, \left(\frac{M}{M_\odot}\right)^{-1/2} P[\log_{10} (M/M_\odot)] \beta \ ,
\eea
where $\beta$ denotes the total PBH fraction, and $P[\log_{10} (M/M_\odot)]\beta$ is the PBH fraction in one logarithmic mass bin. Black holes formed at different times undergo different amounts of dilution due to spatial expansion: the factor $(M/M_\odot)^{-1/2}$ takes this into account. Compared to the distribution $P[\log_{10} (M/M_\odot)]$ without this factor (shown in figure \ref{fig:astmass}), the dilution enhances the contribution of lower masses, as they enter the Hubble radius earlier and thus their energy density is enhanced more relative to radiation. For simplicity, we neglect the $\Cmax$ dependence of the mass \re{mass} and the critical scaling \re{criticalmass} when computing $\Omega_{\text{PBH}}$, so that black holes with the same mass form at the same time. It would be straightforward to include these effects in our analysis pipeline.

In \tab{tab:numbers} we show the monochromatic mass, the 95\% stochastic mass range with and without critical collapse, as well as $\Omega_{\text{PBH}}$ calculated with the full stochastic and the mean profiles with various collapse thresholds. (As we do not include the $\c=0.2$ efficiency factor for the mass, some numbers are slightly different from those in \cite{Figueroa:2021zah}, where further details, including values of CMB observables, can be found.) Including the integral \re{omega_total} over the mass distribution instead of assuming a single PBH mass suppresses $\Omega_{\text{PBH}}$ by two orders of magnitude for $\Cth=0.4$ and less than one order of magnitude for $\avCth=0.4$.

\begin{table}[t]
\begin{adjustwidth}{-1cm}{-1cm}
\begin{center}
\begin{tabular}{lccc}
\toprule
& Asteroid & Solar & Supermassive \\
\midrule
\textbf{Mass (in units of $M_\odot$)} &&& \\
monochromatic & $4\times10^{-14}$ & $20$ & $9\times10^3$ \\
95\% range & $9 \times 10^{-12}\ldots1 \times 10^{-6}$ & $300\ldots1 \times 10^{8}$ & $2 \times 10^{5}\ldots7 \times 10^{10}$ \\
95\% range with critical collapse & $5 \times 10^{-12}\ldots2 \times 10^{-7}$ & $400\ldots1 \times 10^{8}$ & $4 \times 10^{5}\ldots7 \times 10^{9}$ \\
\midrule
\textbf{Monochromatic $\Omega_\text{PBH}$} &&& \\
Gaussian $\zetath=1$ & 0.07 & 0.08 & $7\times10^{-6}$ \\
Gaussian mean $\Cth=0.4$ & 2 & $2\times10^{-9}$ & $2\times10^{-22}$ \\
Gaussian mean $\Cth=0.5$ & $1\times10^{-11}$ & $8\times10^{-22}$ & $7\times10^{-50}$ \\
Gaussian mean $\avCth=0.4$ & $4\times10^{-3}$ & $2\times10^{-10}$ & $3\times10^{-25}$ \\
non-Gaussian mean $\Cth=0.4$ & $9\times10^{3}$ & $2\times10^{-7}$ & $4\times10^{-18}$ \\
non-Gaussian mean $\Cth=0.5$ & $4\times10^{-3}$ & $6\times10^{-17}$ & $5\times10^{-39}$ \\
non-Gaussian mean $\avCth=0.4$ & $300$ & $4\times10^{-8}$ & $7\times10^{-20}$ \\
stochastic $\zetath=1$ & $4\times10^3$ & 0.8 & $1\times10^{-3}$ \\
stochastic $\Cth=0.4$ & $9\times10^{12}$ & $2\times10^{7}$ & $3\times10^{10}$ \\
stochastic $\Cth=0.5$ & $5\times10^{10}$ & $4\times10^{6}$ & $2\times10^{8}$ \\
stochastic $\avCth=0.4$ & $5\times10^{6}$ & $3$ & - \\
\textbf{Full $\Omega_\text{PBH}$} &&& \\
stochastic $\Cth=0.4$ & $9\times10^{10}$ & $8\times10^{5}$ & $1\times10^{9}$ \\
stochastic $\avCth=0.4$ & $1\times10^{6}$ & $600$ & - \\
\midrule
\bottomrule
\end{tabular}
\end{center}
\end{adjustwidth}
\caption{PBH mass and energy density. Top row shows the monochromatic mass for which the potentials were originally tuned in \cite{Figueroa:2021zah}. The second row shows the 95\% range for the mass derived in the present work from the distribution of $\rmax$ with the collapse threshold $\Cth = 0.4$. The third row shows the same quantity with the effects of critical collapse included. We caution that the mass range depends on the resolution of our simulations, and extends to larger values with higher resolution. The next rows show the PBH density parameter today, assuming all PBHs have the mass given in the top row and using \re{omega_M}, determined either for the Gaussian mean profile (equation \re{zeta_mean} with $\epsilon_2=0$ -- the asteroid results differ slightly from \paperI\ due to increased resolution), the non-Gaussian mean profile \re{zeta_mean}, or the stochastic results, and using either $\zeta_{<\kold}$ (see dicussion at the beginning of \ref{sec:ast_pars}), $\Cmax$, or $\avCmax$ as the collapse threshold. After that, we give $\Omega_\text{PBH}$ for the stochastic profiles in the case when they are integrated over the mass distribution as in \re{omega_total}. (In the supermassive case, the $\Omega_\text{PBH}$ values include the enhancement factor of $\approx 5.5\times10^5$, but the masses are the initial masses at the formation time.)
}
\label{tab:numbers}
\end{table}

It has to be checked with collapse simulations whether the peaks at large $\rmax$ that correspond to large masses actually lead to PBH formation. Individual narrow peaks are more likely to be washed away by pressure gradients (which involve the third derivative of $\zeta$, and are thus even more spiky than $\C$) before collapsing than wide peaks or densely packed peaks. High peaks are more likely to be narrow, so many of them may be wiped away, but this can happen at any radius, and on average, the peak width $w$ grows with growing $\rmax$.

While we have not included the conventional $\c=0.2$ collapse efficiency factor for the PBH mass, we have also not taken into account accretion, which for smooth profiles has been found to increase the mass by around one order of magnitude or less \cite{Escriva:2019nsa, Escriva:2021pmf, Escriva:2023qnq, Jangra:2024sif}. Large pressure gradients make accretion less effective, so the evolution can be different for our stochastic profiles, depending on how quickly the spikes are smoothed. The stochastic fluctuations may also enhance asphericity and rotation, which can impact the accretion dynamics. The mass function observed today can also be changed by mergers, which are dependent on the clustering properties and expected to be important for higher mass PBHs. The extended mass function produced by stochastic effects can have a large impact on clustering and mergers, including by producing many mergers with very unequal masses.

\subsubsection{Impact on observational constraints}

If the PBH mass distribution has significant support over a range of masses wider than the four orders of magnitude of the observational asteroid-mass window, it could be already ruled out, or at least the constraints could be much tighter than in the monochromatic case. It has been argued that if all dark matter consists of asteroid-mass PBHs, then the gravitational waves generated at second order by the large scalar perturbations that lead to PBH formation will be detected by the LISA observatory \cite{Cai:2018dig, Bartolo:2018evs, Bartolo:2018rku, Inui:2024fgk, Iovino:2024sgs} (see also \cite{Riccardi:2021rlf, Perna:2024ehx}). However, the analyses of detectability have so far neglected the impact of stochastic effects on the compaction function. In such a simplified setup, the amplitude of the power spectrum of $\zeta$ (together with possible non-Gaussianity) determines both the PBH abundance and the gravitational wave amplitude, but we have found that $\zetamax$ is not strongly correlated with $\Cmax$ and hence PBH abundance. Also, the non-Gaussianity generated by stochastic effects is significant only in the tail (like the non-Gaussianity in \cite{Inui:2024fgk} but unlike the non-Gaussianity in \cite{Cai:2018dig}): the mean of the distribution, where most gravitational waves are generated, remains Gaussian. Overall, when stochastic effects and the compaction function are considered, the observed dark matter abundance is obtained with a smaller amplitude of the power spectrum, leading to weaker gravitational waves. Quantifying these changes would require going much further into the tail of the distribution; we present some estimates in \sec{sec:disc}. The frequency of gravitational waves is also affected by the change in $\kpeak$: as stochastic effects push the mass distribution to higher masses, $\kpeak$ has to be increased to compensate, shifting the gravitational wave signal to higher frequencies.

It has been proposed to detect asteroid-mass PBHs via their influence on Solar system and Earth orbits \cite{Tran:2023jci, Thoss:2024vae}, and constraints have recently been put in the asteroid mass range using the stellar mass function of ultra-faint dwarf galaxies \cite{Esser:2025pnt}. These constraints and searches depend on the PBH mass distribution, on which the stochastic effects can have a large impact.

\subsection{Solar mass dark matter} \label{sec:sol}

\subsubsection{Parameters}

There are strong constraints against PBHs in the solar-mass range being all of the dark matter, including from microlensing, X-rays from accretion, and gravitational waves from PBH binaries \cite{Carr:2020gox, Green:2020jor}. It has been argued that some of the constraints could be evaded by clustering, but this appears rather uncertain. Even if solar-mass PBHs are not all of the dark matter, they could form a subdominant component. It has also been suggested that some of the gravitational wave events observed by the LIGO/Virgo/KAGRA collaboration are due to solar-mass PBHs instead of black holes of stellar origin. So far, the observed masses and other properties do not seem to be in contradiction with the astrophysical understanding of the origin and evolution of stellar black holes, but observation of sub-solar black hole mergers would be strong evidence for PBHs \cite{Rantala:2024crf}. It has also been proposed that the pulsar timing array evidence for gravitational waves could result from second order scalar perturbations related to solar-mass PBHs \cite{Kalaja:2019uju, Gow:2020bzo, Unal:2020mts, Franciolini:2023pbf, Inomata:2023zup, Wang:2023ost, Liu:2023ymk, Iovino:2024tyg, Inui:2024fgk}. Aside from observational considerations, the solar case in comparison to the asteroid case helps to understand the dependence of the stochastic effects on the details of the potential and the resulting power spectrum.

Solar-mass PBHs form around the time of the QCD crossover, when the equation of state is softened from $\frac{1}{3}$, enhancing PBH production: we do not take this into account. Then the only difference between the solar and the asteroid case is the potential, which leads to $\epsilon_2=0.287$ and a wider power spectrum than in the asteroid case, with $\kpeak=1.3\times10^5$ Mpc$^{-1}$ and $\mathcal P_\zeta(\kpeak)=1.2\times10^{-2}$. Similarly to the asteroid case, the potential was originally tuned to give (with the $\c=0.2$ efficiency factor) $M=4.7 M_\odot$ and $\Omega_\text{PBH}=0.17$ when stochastic effects are neglected and the collapse threshold is taken to be $\zeta=1$ with Gaussian statistics \cite{Figueroa:2020jkf, Figueroa:2021zah}. Without the $\c$ factor, the mass is 23 $M_\odot$, and the mass corresponding to $\kpeak$ is 870 $M_\odot$. 

We reach up to $\Cmax=2/3$ and $\avCmax=0.42$, higher than in the asteroid case. Unlike in the asteroid case, we have 17 simulations (out of $10^8$) with $r\zeta'<-1$, resulting in type II profiles with $\Cmax$ exactly equal to $2/3$ \cite{Kopp:2010sh, Harada:2013epa, Carr:2014pga, Uehara:2024yyp, Harada:2024jxl, Shimada:2024eec}. It is not clear what is the mass of PBHs forming from type II profiles. We assume that just as in the case of type I profiles, it is simply the mass inside $\rmax$. As the fraction of type II profiles is small, their presence has negligible effect on our results. (For type II profiles, $\avCmax$ is not defined because $R$ is not monotonic.) The number of peaks $p$ of $\C(r)$ above $\Cth=0.4$ follows the scaling law $n(p)\propto e^{-\a p}$ with $\a=1.0$.

\subsubsection{Abundance}

\begin{figure}[th]
  \centering
  \includegraphics[scale=1]{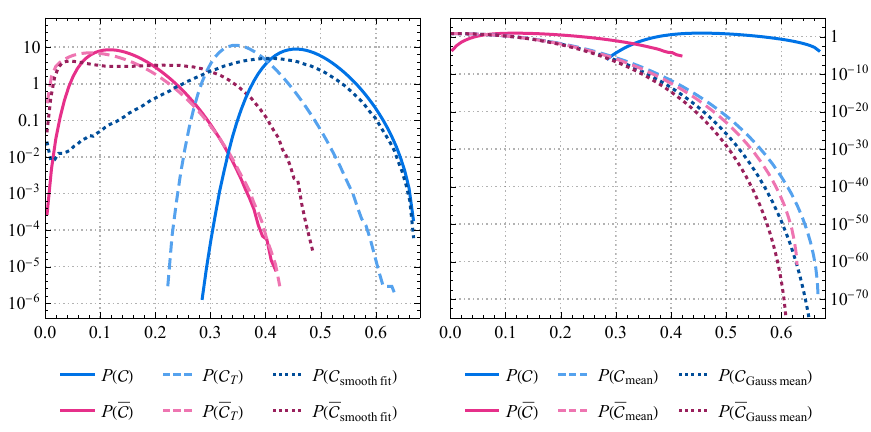}
  \caption{The probability distributions of $\Cmax$ and $\avCmax$ in the solar case, similar to \fig{fig:astP}.}
  \label{fig:solarP}
\end{figure}

In \fig{fig:solarP} we show the probability distributions of $\Cmax$ and $\avCmax$ for the stochastic case, with and without the transfer function. We also show the probability distribution inferred from the smooth fit profiles and from the full mean and Gaussian mean profiles.

For $\Cth=0.4$ and the full stochastic profiles, we have $\b=0.9$ -- practically all matter collapses into PBHs. (Here the approximation of independent stochastic patches fails.) Raising the threshold to $\Cth=0.5$ has little effect on the abundance, lowering it to $\b=0.2$. Using $\avCth=0.4$ instead gives $\b=2\times10^{-7}$. The decrease from $\Cth=0.4$ to $\Cth=0.5$ is somewhat larger than in the asteroid case, while the ratio between the $\Cth=0.4$ and the $\avCth=0.4$ abundance is almost the same as in the asteroid case.

The left plot of \fig{fig:solarP} shows that, as in the asteroid case, the transfer function \re{T} has a much more pronounced effect on $\Cmax$ than on $\avCmax$: for large values, the distribution of $\avCmax$ is hardly affected. Including the transfer function reduces $\b$ by one order of magnitude for $\Cth=0.4$ and two orders of magnitude for $\Cth=0.5$, both an order of magnitude less than in the asteroid case. In the solar case there are enough realisations to also evaluate the effect for $\avCth=0.4$: $\b$ falls by less than a factor of 2, although the precise number is probably not well resolved, as the number of realisations above the threshold is small. As in the asteroid case, the distribution of $\avCmax$ determined from the smooth fit has roughly the right slope for large values, but the normalisation is off, here by three orders of magnitude. Also similarly to the asteroid case, the distribution of $\Cmax$ from the smooth fit asymptotes to the real distribution for large values.

In the right plot of \fig{fig:solarP}, we again compare the full $\Cmax$ and $\avCmax$ distributions to the ones for the non-Gaussian and Gaussian mean profiles. The behaviour is similar to the asteroid case. In the Gaussian case, using the mean profile gives $\b=1\times10^{-16}$ for $\Cth=0.4$, $\b=4\times10^{-29}$ for $\Cth=0.5$, and $\b=8\times10^{-18}$ for $\avCth=0.4$. As in the asteroid case, the abundance calculated from the Gaussian mean profile is both much smaller than the abundance from the stochastic calculation and falls much more steeply with rising collapse threshold. The original Gaussian result with the threshold $\zeta_{<\kold}>\zetath=1$ gives $\b=4\times10^{-9}$, and including the non-Gaussian tail gives $\b=4\times10^{-8}$. Unlike in the asteroid case, for Gaussian perturbations using $\Cth$ instead of $\zetath$ suppresses the abundance instead of enhancing it. For $\Cth=0.4$, the full stochastic enhancement over the Gaussian case is a factor of $10^9$ while the effect of including the exponential tail in the distribution of $\zeta$ gives only a factor of $10$. The enhancement is four orders of magnitude smaller than in the asteroid case, likely because of the smaller value of $\epsilon_2$, which controls the amplitude of the non-Gaussianity.

We do not show correlation plots, because they are very similar to the asteroid case: $\Cmax$ and $\avCmax$ are rather uncorrelated, $\zetamax$ does not predict $\Cmax$ and is better correlated with $\avCmax$. 

\begin{figure}[th]
  \centering
  \includegraphics[scale=1]{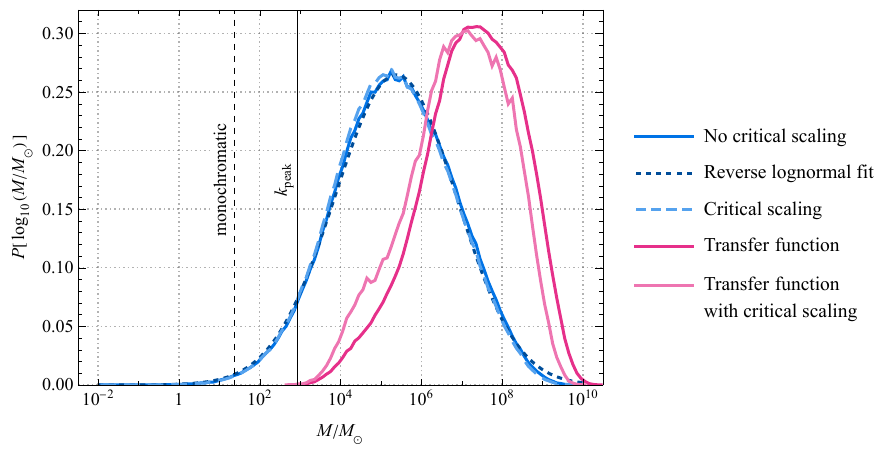}
  \caption{The mass distribution in the solar case, similar to \fig{fig:astmass}. The vertical lines mark the monochromatic mass considered in \cite{Figueroa:2021zah} and the mass corresponding to $\kpeak$, at 23 $M_\odot$ and 870 $M_\odot$, respectively.}
  \label{fig:solarmass}
\end{figure}

\subsubsection{Mass distribution}

In \fig{fig:solarmass} we show the mass distribution, calculated assuming $\Cth=0.4$, both with and without the effect of critical collapse \re{criticalmass} and with and without the transfer function \re{T}. The reverse lognormal function in the variable $\log_{10} (M/M_\odot)$ is a better fit to the distribution than in the asteroid case, although it again overpredicts the high-mass tail. The 95\% mass range without the critical scaling or the transfer function is $300\ldots1\times10^8 M_\odot$, well above the mass that the model was originally constructed for, as in the asteroid case. Including the effects of critical collapse has a smaller effect than in the asteroid case, and the 95\% mass range is $400\ldots1\times10^8 M_\odot$ for both the narrow and extended range of critical scaling. Including the transfer function again moves the distribution up without significantly affecting its width, and the 95\% mass range becomes $4\times10^3\ldots2\times10^9 M_\odot$. Including critical scaling with the transfer function shifts this to somewhat smaller masses, but not as much as in the asteroid case. As in the asteroid case, the mass distribution has not converged, and the peak shifts to higher masses and widens with increasing resolution. Including the integral \re{omega_total} over the mass distribution suppresses $\Omega_{\text{PBH}}$ by one to two orders of magnitude.

\subsubsection{Impact on observational constraints} \label{sec:solarobs}

The amplitude of the possibly detected pulsar timing array signal is observationally fixed and, if produced by the second order effect of scalar perturbations, directly related to the amplitude of the power spectrum. The resulting amplitude can easily overproduce PBHs even without stochastic effects \cite{Kalaja:2019uju, Gow:2020bzo, Unal:2020mts, Franciolini:2023pbf, Inomata:2023zup, Wang:2023ost, Liu:2023ymk, Iovino:2024tyg, Inui:2024fgk}. Any stochastic enhancement makes this explanation even more difficult, although the related non-Gaussianity can loosen the constraints \cite{Nakama:2016gzw}. The stochastic widening of the mass distribution means that solar-mass PBHs would be accompanied by PBHs in the range that has been considered as the seeds of supermassive black holes, as we see from \fig{fig:solarmass}. Explaining both the pulsar timing array signal and supermassive seeds at the same time might thus be natural, although the wider mass range also brings in new observational constraints, including spectral distortions, which we discuss in \sec{sec:superobs}.

\subsection{Supermassive black hole seeds} \label{sec:seed}

\subsubsection{Parameters}

To produce seeds for supermassive black holes in the centres of galaxies, we have $\kpeak=5.4\times10^3$ Mpc$^{-1}$, with the value $\mathcal P_\zeta(\kpeak)=5.3\times10^{-3}$. The shape of the power spectrum and the value $\epsilon_2=0.294$ are close to those of the solar case, but the peak value of the power spectrum is smaller. The potential was tuned to give PBH mass (with $\c=0.2$) $1.8\times10^3 M_\odot$. Without the $\c$ factor the mass is $9.1\times10^3 M_\odot$, and the mass corresponding to $\kpeak$ is $4.7\times10^5 M_\odot$. In \cite{Figueroa:2021zah} we assumed that accretion brings the mass up to $10^9 M_\odot$, and the density parameter today is $\Omega_\text{PBH}=1.4\times10^{-5}$, roughly corresponding to the mass density of observed central supermassive black holes. In table \ref{tab:numbers} we assume the same enhancement factor of $\approx5.5\times10^5$ for the density parameter due to accretion. (All mass values we give refer to the initial mass, without the enhancement factor. It is included only in the values of $\Omega_\text{PBH}$ shown in table \ref{tab:numbers}.) We reach up to $\Cmax=0.60$ and $\avCmax=0.34$. The number of peaks $p$ of $\C(r)$ above $\Cth=0.4$ follows the scaling law $n(p)\propto e^{-\a p}$ with $\a=1.7$.

\begin{figure}[th]
  \centering
  \includegraphics[scale=1]{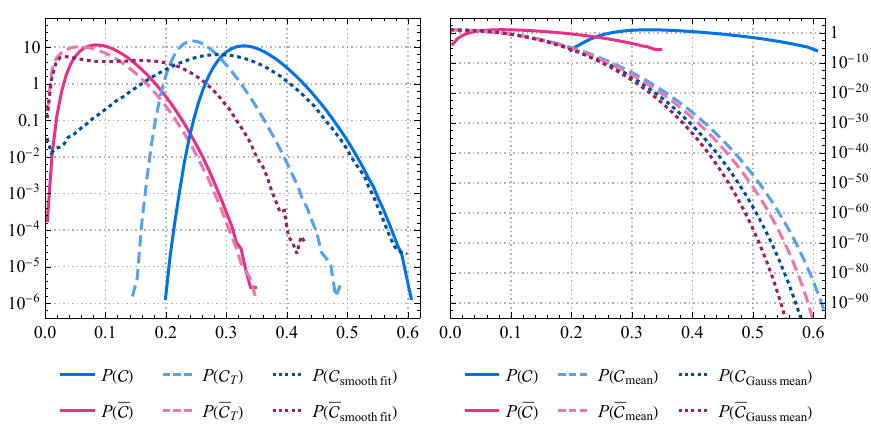}
  \caption{The probability distributions of $\Cmax$ and $\avCmax$ in the supermassive case, similar to \fig{fig:astP}.}
  \label{fig:superP}
\end{figure}

\subsubsection{Abundance}

On the left of \fig{fig:superP} we show the probability distributions of $\Cmax$ and $\avCmax$ for the stochastic case, with and without the transfer function. We also show the probability distribution inferred from the smooth fit profiles.

For the full stochastic case, we have $\b = 0.05$ for $\Cth=0.4$, and $\b=4\times10^{-4}$ for $\Cth=0.5$. As in the asteroid case, the precise number may suffer from resolution effects, but the order of magnitude has converged; see appendix \ref{app:res} for details. We do not have enough realisations to reach $\avC=0.4$.

Again, the transfer function \re{T} has a large effect on the distribution of $\Cmax$ and a small effect on the distribution of $\avCmax$. The impact is bigger than in the asteroid and the solar case: including the transfer function reduces $\b$ by three orders of magnitude for $\Cth=0.4$. With the transfer function, there are no realisations where $\Cmax>0.5$, indicating a suppression of at least four orders of magnitude. The distributions of $\avCmax$ and $\Cmax$ determined from the smooth fits again show the same features for large values: for $\avCmax$, the right slope but wrong normalisation, and for $\Cmax$, asymptotic approach to the real distribution (the deviation at the end of the tail is again likely a numerical artifact).

On the right of \fig{fig:superP} we again show the $\Cmax$ and $\avCmax$ distributions derived in the full stochastic case and with the non-Gaussian and Gaussian mean profiles: the behaviour is similar to the asteroid case. The Gaussian mean profile gives $\b=3\times10^{-34}$ for $\Cth=0.4$, $\b=1\times10^{-61}$ for $\Cth=0.5$, and $\b=6\times10^{-37}$ for $\avCth=0.4$. The Gaussian mean abundances are much smaller than the full stochastic results, even more so than in the asteroid and solar cases. The Gaussian result for $\zetath=1$ is $\b=1\times10^{-17}$, and including the non-Gaussian tail gives $\b=2\times10^{-15}$. As in the solar case, for Gaussian perturbations using $\Cth$ instead of $\zetath$ suppresses the abundance instead of enhancing it. The effect of the spikiness of $\C(r)$ is again several orders of magnitude larger than that of the exponential tail of $\zeta$.

The correlations between $\Cmax$, $\avCmax$, and $\zetamax$ are very similar to the previous cases, so we again omit the plots.

\begin{figure}[th]
  \centering
  \includegraphics[scale=1]{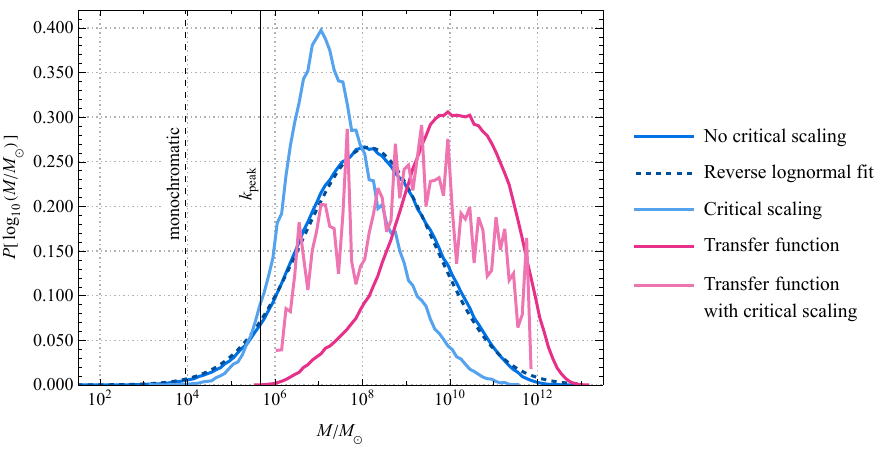}
  \caption{The mass distribution in the supermassive case, similar to \fig{fig:astmass}. The vertical lines mark the monochromatic mass considered in \cite{Figueroa:2021zah} and the mass corresponding to $\kpeak$, at $9.1\times10^3 M_\odot$ and $4.7\times10^5 M_\odot$, respectively.}
  \label{fig:supermass}
\end{figure}

\subsubsection{Mass distribution}

In \fig{fig:supermass} we show the mass distribution, calculated assuming $\Cth=0.4$, both with and without the effect of critical collapse \re{criticalmass} and with and without the transfer function \re{T}. As in the solar case, the reverse lognormal function in the variable $\log_{10} (M/M_\odot)$ is a reasonable fit to the distribution, but again overpredicts the high-mass tail. The 95\% mass range is $2\times10^{5}\ldots7\times10^{10} M_\odot$, again well above the original mass for the model. Including the integral \re{omega_total} over the mass distribution suppresses $\Omega_{\text{PBH}}$ by one order of magnitude. Critical collapse again shifts mass towards smaller values, $4\times10^{5}\ldots7\times10^{9} M_\odot$, except at the lower tail, about as strongly as in the asteroid case. The difference between the narrow and the extended range of applicability for critical scaling is again small. The transfer function again moves the distribution up without much change in its width: the 95\% mass range becomes $2\times10^{7}\ldots9\times10^{11} M_\odot$. Including also critical scaling shifts the mass closer to the case without either transfer function or critical scaling, but the data is rather noisy due to sparse sampling at high masses.

\subsubsection{Impact on observational constraints} \label{sec:superobs}

PBHs as seeds of supermassive black holes at the centres of galaxies are constrained, among other observations, by spectral distortions in the CMB caused by Silk damping of the large density perturbations on PBH scales \cite{Unal:2020mts, Sharma:2024img, Byrnes:2024vjt, Iovino:2024tyg}. For a Gaussian power spectrum (and using the value of $\zeta$ as the collapse criterion), PBHs of mass $10^4 M_\odot$ or above and sufficient number density to act as seeds are excluded. It has been argued that PBHs with mass $10^3 M_\odot$ are too small to act as the seeds of supermassive black holes, at least for the density fraction $10^{-5}$ that we have used, but there is much uncertainty about how supermassive black holes form \cite{Inayoshi:2019fun, Prole:2025snf}. As the spectral distortion constraints are based on the amplitude of the power spectrum, the stochastic enhancement makes them looser (when keeping the PBH abundance fixed), similarly to the upcoming future gravitational wave constraints from the LISA observatory. Using the maximum value of $\C$ instead of $\zeta$ as the collapse criterion has been found to loosen the constraints by five orders of magnitude, even without taking into account non-Gaussianity or stochastic effects \cite{Yang:2024snb}. An extended mass spectrum may also alleviate the constraints, in addition to changing the way PBHs seed heavier black holes. At the same time, an extended mass function may bring multiple types of constraints to bear. In addition to clustering and mergers, migration and ejection from the host galaxy can also affect the observed mass distribution \cite{Siles:2024yym}, and the observed black hole range is likely biased, with black holes in some mass ranges remaining undiscovered \cite{Inayoshi:2019fun}. The pulsar timing array signals discussed in \sec{sec:solarobs} also provide constraints for PBHs above solar mass. These will be stricter in the future, although they depend on the degree of non-Gaussianity and will be alleviated by the lowering of the amplitude of the power spectrum because of the stochastic enhancement \cite{Unal:2020mts, Iovino:2024tyg}.

\section{Discussion} \label{sec:disc}

\para{Spherical symmetry.}

Spherical symmetry is a key technical assumption in our calculation. For a Gaussian field, high peaks (which are needed for PBH formation) are close to spherically symmetric \cite{Bardeen:1985tr} -- although see \cite{Young:2022phe} for criticism of the loose identification of peaks of $\zeta(r)$ and peaks of $\C(r)$. In our situation, it is not clear how good the assumption of spherical symmetry is, and in some cases non-Gaussianity can generate significant asphericity \cite{Germani:2025fkh}. We find that when the maximum of $\C(r)$ is large, $\zeta$ often does have a maximum in the centre, although there is a lot of variation. However, our patches are not selected to be centred on a peak: the origin is a random point in space. The problem may be alleviated by the fact that cases with small mass excess, which we expect to be most likely to violate spherical symmetry, are irrelevant for PBH formation. A lattice calculation of inflation in the stochastic $\Delta N$ formalism has reported violations of spherical symmetry of order $10\%$ \cite{Mizuguchi:2024kbl}, but analytical studies and simulations of non-spherical collapse of a radiation fluid find that non-spherical modes decay and the effects of asphericity are small \cite{Gundlach:1999cw, Kuhnel:2016exn, Celestino:2018ptx, Yoo:2020lmg, Escriva:2024aeo, Escriva:2024lmm, Harada:2024jxl, Yoo:2024lhp}. The fact that the spherically symmetric compaction functions we obtain are far from smooth suggests that the large stochastic fluctuations could also lead to significant violations of spherical symmetry. The importance of violations of spherical symmetry should be quantified with three-dimensional stochastic calculations of the initial conditions, followed with three-dimensional relativistic collapse simulations. Running collapse simulations with the stochastic profiles in the spherically symmetric case would be a first step in this direction.

\para{Smaller power spectrum.}

The more the stochastic effects increase the abundance, the smaller the amplitude of the inflationary power spectrum on PBH scales has to be to match the observed density of dark matter or the required density of supermassive black hole seeds. The stochastic effects increase the abundance by several orders of magnitude, and the maximum amplitude of the power spectrum in all three cases is $\sim10^{-2}$.

In \cite{Kawasaki:2019mbl, Young:2019yug} it was found that suppression of the abundance by many orders of magnitude due to using the compaction function (with the Gaussian mean profile) instead of $\zeta$ as the collapse criterion could be compensated with an adjustment of a factor of 2 or so in the power spectrum. In \cite{Atal:2019cdz} enhancement due to non-Gaussianities was compensated by reducing the amplitude of the power spectrum by a factor of 10. A similar result was found in \cite{Tomberg:2023kli}. In \cite{Abe:2022xur}, taking into account the exponential tail was found to reduce the amplitude of the power spectrum by a factor of 4, down to $10^{-3}$. In \cite{Wilkins:2023asp} a spectator waterfall field enhanced the stochastic noise term, so the amplitude of the power spectrum could be below $10^{-3}$ and still produce the right abundance, while \cite{Escriva:2023uko} obtained the right PBH abundance for peak amplitude $5\times10^{-4}$ due to contribution from type II PBH profiles generated by the field getting stuck in the false vacuum. In \cite{Inui:2024fgk} the effect of phenomenological non-Gaussianity was considered, and it was found that the amplitude of the power spectrum (taken to be a Dirac delta function) could be as small as $7\times10^{-4}$ while getting the observed abundance of PBHs as dark matter. For details on the relation between the power spectrum and the PBH abundance in the Gaussian case, including discussion of the effects of critical collapse, asphericity and dependence on profile shape, see \cite{Akrami:2016vrq, Kalaja:2019uju, Gow:2020bzo}.

In our stochastic case, it is not straightforward to map the amplitude of the power spectrum to the abundance, nor its shape to the mass distribution. This is reflected in the fact that the amplitude of $\Cmax$ is not predicted by the amplitude of $\zetamax$ (as shown in figure \ref{fig:astcor}). Physically, the property that the amplitude of the power spectrum could be much smaller than required when using $\zeta$ as the criterion corresponds to the fact that the gradients of a rapidly varying function can be large regardless of its amplitude -- though here the amplitude of the stochastic kicks is also controlled by the power spectrum. If the power spectrum is much smaller, the potential can be less tuned, possibly quantitatively slightly alleviating the fine-tuning problems of inflationary PBH models \cite{Cole:2023wyx}. The other possibility, which seems more likely, is that the power spectrum doesn't need to change much, so the impact of stochasticity can be compensated by a slight adjustment in the potential, in which case uncertainty due to stochastic effects is swamped by lack of knowledge about the details of the potential.

\begin{figure}[th]
  \centering
  \includegraphics{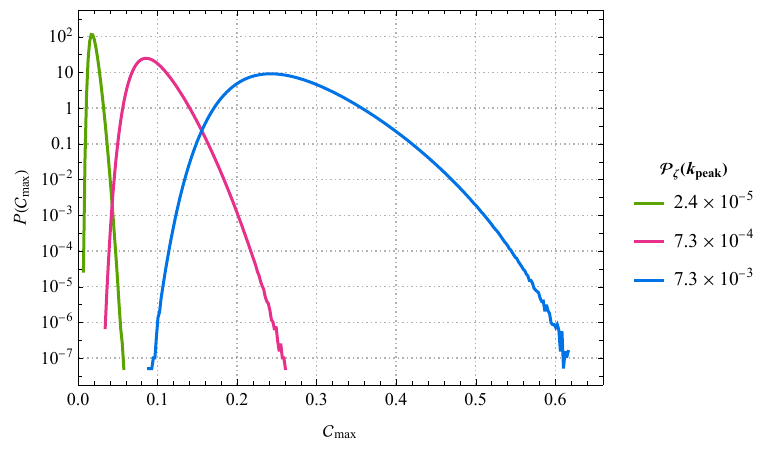}
  \caption{The probability distributions of $\Cmax$ in the asteroid case for various amplitudes of the power spectrum.}
  \label{fig:varied_PR}
\end{figure}

To test these possibilities, we ran new asteroid mass simulations with two potentials modified to produce smaller power spectra in the USR region, while still agreeing with CMB observations. The models have $\Pzeta(\kpeak)=7.3\times10^{-4}$, $\epsilon_2=0.677$ and $\Pzeta(\kpeak)=2.4\times10^{-5}$, $\epsilon_2=0.470$. Compared to our original asteroid mass model with $\Pzeta(\kpeak)=7.3\times10^{-3}$, $\epsilon_2=0.807$, the power spectra are smaller by the factors $0.1$ and $3\times10^{-3}$, and the $\epsilon_2$ values are also lower.\footnote{The values of $\Pzeta(\kpeak)$ and $\epsilon_2$ are correlated, since $\epsilon_2$ sets the rate at which $\Pzeta(k)$ decreases from its maximum towards its (approximately fixed) end-of-inflation value \cite{Karam:2022nym}.} We ran $10^{10}$ simulations of each case (with resolution $\rmd\ln k = 1/32$ and $\rmd\ln r = 0.01$; see appendix \ref{app:res} for discussion of the resolution). The model with $\Pzeta(\kpeak)=7.3\times10^{-4}$ reached the maximum value $\Cmax = 0.27$, while the model with $\Pzeta(\kpeak)=2.4\times10^{-5}$ only reached $\Cmax = 0.06$. The base case had a maximum of $\Cmax = 0.63$. In \fig{fig:varied_PR} we plot the resulting $\Cmax$ distributions along with the distribution corresponding to the original value of the power spectrum. These simulations have lower resolution than those whose results are shown in \fig{fig:astP}, and we have not conducted similar extensive convergence checks for them, as the results are in any case only indicative. Reducing $\Pzeta(\kpeak)$ strongly suppresses the values of $\Cmax$. We are far from having enough realisations to reach $\Cmax=0.4$, but a naive extrapolation of the distributions to larger values suggests that $\Pzeta$ has to be decreased by less than two orders of magnitude to reach the observed dark matter abundance, which corresponds to $\beta\sim 10^{-17}\ldots10^{-15}$.

These estimates are preliminary. Further study is needed to disentangle the effects of $\Pzeta(\kpeak)$ and $\epsilon_2$ and build distributions that reach $\Cth=0.4$ and give a realistic dark matter abundance. If the distribution is sampled fairly, the number of required realisations grows rapidly as PBHs become rarer -- naively to the factor of $10^{15}\ldots10^{17}$. Even with our analytical approximation \re{phi_solution}--\re{zeta_vs_X}, this is unfeasible, as the vast majority of the computational time is wasted on irrelevant patches where $\Cmax$ is small. Instead, methods that allow calculating only the tail of the distribution relevant for PBH formation or at least sampling it more efficiently are required, such as importance sampling \cite{Jackson:2022unc, Tomberg:2022mkt, Mizuguchi:2024kbl, Jackson:2024aoo} or stochastic trees \cite{Animali:2025pyf}.

Even if keeping the PBH abundance the same involves only a slight retuning of the potential, this can have a large effect on the relation between the PBH abundance and the amplitude of gravitational waves, relevant for LISA and pulsar timing array observations, as we have noted above. Furthermore, the effect on the mass distribution likely cannot be undone by modifying the potential (as discussed, this will also affect the frequency of the second order gravitational waves).

The amplitude of the tree-level power spectrum can also be important for one-loop corrections. It has been argued that if $\mathcal P_\zeta\sim10^{-2}$ on PBH scales, perturbation theory fails on CMB scales, invalidating all studies of PBH generation in single-field inflation \cite{Inomata:2022yte, Kristiano:2022maq}. This claim has been the subject of intense debate, and the most recent and comprehensive studies find that there is no such effect from small-scale modes on the large-scale modes \cite{Inomata:2022yte, Kristiano:2022maq, Riotto:2023hoz, Choudhury:2023vuj, Kristiano:2023scm, Riotto:2023gpm, Firouzjahi:2023aum, Motohashi:2023syh, Firouzjahi:2023ahg, Franciolini:2023agm, Tasinato:2023ukp, Cheng:2023ikq, Fumagalli:2025qie, Maity:2023qzw, Tada:2023rgp, Firouzjahi:2023bkt, Davies:2023hhn, Iacconi:2023ggt, Inomata:2024lud, Ballesteros:2024zdp, Kristiano:2024vst, Kristiano:2024ngc, Kawaguchi:2024rsv, Fumagalli:2024jzz, Kong:2024lac, Sheikhahmadi:2024peu, Inomata:2025bqw, Fang:2025vhi, Inomata:2025pqa, Braglia:2025qrb, Kristiano:2025ajj}. If such an effect did exist, stochastic enhancement could save perturbation theory by decreasing the amplitude of the power spectrum by one or two orders of magnitude, as studied in the above example.

\para{Importance of non-Gaussianity.}

In our approximation, non-Gaussianity is purely controlled by the slow-roll parameter $\epsilon_2$, which affects the profile \re{zeta_2} (and its modified version \re{r_zeta_prime_2}). Non-Gaussianity is important if the quantity $1-\frac{\epsilon_2}{2}X_{<k}$ is clearly different from one. In all of our stochastic profiles, $1-\frac{\epsilon_2}{2}X_{<k}$ stays positive, but for high values of $X_{<k}$, it can reach values of order $0.1$, changing the compaction function by an order one factor. At linear order, $X_{<k} \approx \zeta_{<k}$ (as is clear from \re{zeta_vs_X}) so, given that $\epsilon_2 \sim 0.1\dots 1$ in our examples, non-Gaussianity becomes important when $\zeta$ approaches one. As discussed above, $\zeta$ is not strongly correlated with $\C$, so among our PBH-forming profiles there are both cases that are close to Gaussian and cases that deviate from Gaussianity. Profiles with large values of $\C$ typically have $\zeta, X>0$, so non-Gaussianity further enhances $\C$ and thus increases PBH abundance. Since PBHs arise from rare large fluctuations, an order one correction to $\C$ changes the predicted PBH abundance by many orders of magnitude. We have not performed full stochastic simulations without the $\epsilon_2$ parameter, but for the mean profiles \re{zeta_mean} and \re{r_zeta_prime_mean}, putting $\epsilon_2=0$ decreases the predicted abundance by 2 to 5 orders of magnitude, as shown in \tab{tab:numbers}. Non-Gaussianity may be important not only for PBH abundance, but also other statistical properties such as mass distribution and clustering.

\section{Conclusions} \label{sec:conc}

Following up on \paperI, we have evaluated the effect of stochastic kicks during USR on the PBH abundance and mass function when the PBH formation threshold is set by the maximum value of the compaction function $\C(r)$ or its average. We start from inflaton potentials that agree with CMB observations, and aim to minimise the number of ad hoc assumptions. Using approximations that allow us to obtain an analytical solution to the Langevin equation in the relevant regime in terms of Gaussian random variables, we have simulated $10^8$ spherically symmetric patches where PBHs may form, following $4 \times 10^4$ momentum shells in each to find $\C(r)$. We do not introduce any scales beyond those provided by the power spectrum by hand, and do not use any window function for PBH formation. We find all nested peaks, which helps in solving the peak-in-a-peak problem, given a prescription for which maxima form PBHs. We take the PBH mass to be determined stochastically by the scale where $\C(r)$ (or its average) reaches its maximum. We have separately considered PBHs in three cases: asteroid mass range, solar mass range, and $\sim10^3 M_\odot$ relevant for supermassive black hole seeds.

It is well established that stochastic effects enhance PBH abundance by generating an exponential tail for the probability distribution of the curvature perturbation $\zeta$ \cite{Pattison:2017mbe, Ezquiaga:2019ftu, Figueroa:2020jkf, Pattison:2021oen, Figueroa:2021zah, Tomberg:2021xxv, Cai:2022erk, Tomberg:2022mkt, Gow:2022jfb, Tomberg:2023kli, Vennin:2024yzl, Inui:2024sce, Sharma:2024fbr, Animali:2024jiz, Launay:2024qsm}. We have shown that the enhancement due to the increase of $\C(r)$ by the stochastic kicks is even larger. In the supermassive case the abundance is enhanced by 32 orders of magnitude over the Gaussian mean when using the compaction function, and at least 36 orders of magnitude when using the average compaction function, but these values vary a lot depending on the mass case and the collapse threshold. The enhancement is easy to understand, given that $\C(r)$ depends on the first derivative of $\zeta(r)$, and derivatives are more sensitive to small-scale noise. The PBH mass range is also affected: the mass distribution shifts to larger masses and becomes wider. The resulting PBH mass function is much more extended than usually considered in the literature: in all three mass cases, the 95\% mass range spans at least five orders of magnitude. In order to get the right abundance, the amplitude of the power spectrum has to be decreased and the scale $\kpeak$ of its maximum shifted to larger values. Both effects can have a significant impact on the observational constraints on PBHs, including from CMB spectral distortions and from gravitational waves probed by pulsar timing arrays and the LISA observatory. We have also considered the effect of critical collapse, which generally shifts the distribution to smaller masses, and found that its impact is much smaller than the widening due to the stochastic kicks. Including the linear radiation era transfer function has a larger effect than critical scaling. It erases small-scale structure, and thus shifts the distribution to higher masses, but without significantly affecting its width. The transfer function modulates, but does not counter, the stochastic widening of the mass function.

We were not able to reach numerical convergence on the mass distribution. With increasing resolution, the mass function continues to get wider and the mean mass shifts to larger values. So while this change in the mass distribution is robust, we cannot fully quantify it, and determining the resulting change in the gravitational wave spectrum requires further work. Furthermore, and affecting also the abundance (for which we have reasonable convergence) we use collapse thresholds obtained from simulations of smooth profiles, but the real thresholds may be quite different for the spiky stochastic profiles we consider, given that pressure gradients resist collapse. At the same time, we expect that some spikes will be smoothed out by pressure gradients during sub-Hubble evolution before collapse, implying that using the maximum value of the spiky stochastic profiles as the collapse criterion may overestimate the impact of stochastic effects. We have tested this by using the linear radiation era transfer function that smooths small-scale structure before the large-scale spikes enter the Hubble radius. This reduces the stochastic enhancement by a factor that varies from order unity to more than four orders of magnitude, depending on the collapse threshold, but does not eliminate it. As the results for the abundance change by many orders of magnitude when varying the collapse threshold, which is known to depend on the profile shape, our quantitative findings of the abundance can only be considered preliminary.

The collapse threshold has to be checked by redoing collapse simulations for the stochastic profiles, which we are currently working on. Furthermore, the spherical symmetry approximation should be checked with three-dimensional stochastic calculations, and possibly eventually with three-dimensional collapse simulations. The many orders of magnitude change we find and the strong dependence on the collapse criterion show that all current estimates of PBH abundance and mass in the literature have to considered preliminary until the impact of the spikes on the collapse process has been properly evaluated. It remains to be determined how much smaller the power spectrum can be while explaining dark matter or supermassive black hole seeds with PBHs. As the initial realistic PBH abundance is extremely small, this requires methods that focus on the tail instead of sampling the distribution fairly. These steps are key for building a reliable bridge from the inflaton potential to the PBH abundance and mass function.

\acknowledgments

SR thanks Cristiano Germani for useful discussions and Laboratoire de Physique de l’Ecole Normale Sup\'erieure for hospitality. ET was supported by the Lancaster–Sheffield Consortium for Fundamental Physics under STFC grant: ST/X000621/1 and by the ``Fonds de la Recherche Scientifique'' (FNRS) under the IISN grant number 4.4517.08. The authors acknowledge CSC -- IT Center for Science, Finland, for computational resources.

\appendix

\section{Resolution and convergence} \label{app:res}

\begin{figure}[th]
  \centering
  \includegraphics{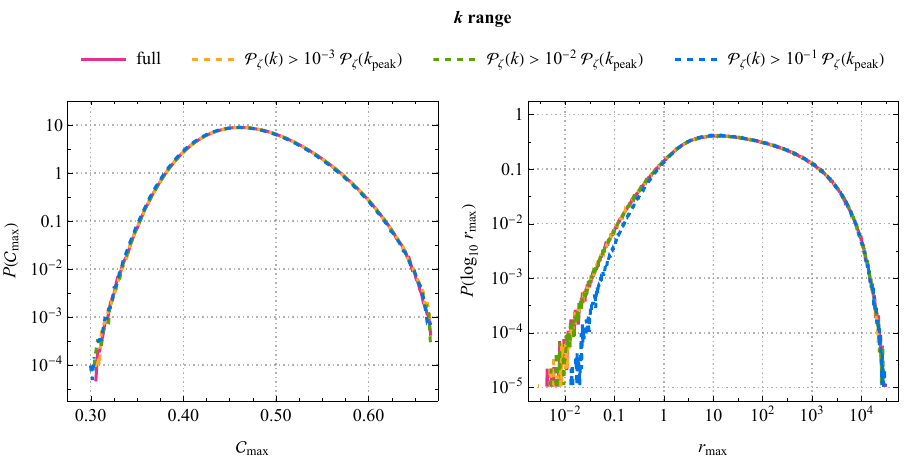}
  \caption{The probability distributions of $\Cmax$ and $\rmax$ with fixed resolutions $\dd \ln k = 10^{-4}$ and $\dd \ln r = 2\times10^{-2}$, but varying $\kini$ and $\kend$. The values $\kini$ and $\kend$ were chosen so that $\Pzeta(\kini)$ and $\Pzeta(\kend)$ equalled a fixed fraction of the maximum value $\Pzeta(\kpeak)$.}
  \label{fig:solar_kend}
\end{figure}

To resolve the compaction function accurately, we need sufficient resolution both in the wavenumber $k$ (when computing the sums \re{zeta_2} and \re{r_zeta_prime}) and in the radial coordinate $r$ (when sampling the radial profiles such as those shown in \fig{fig:astprof}). We use a logarithmic grid for both $k$ and $r$, with step lengths $\dd \ln k$ and $\dd \ln r$, respectively. The resolution is high enough if the summands in \re{zeta_2} and \re{r_zeta_prime} change little between two points (disregarding the random kicks $\hxi_k$). As long as $\dd\ln k \ll 1$, this is easily satisfied for the power spectrum $\Pzeta(k)$, but the integration kernels involving the product $kr$ require more care. In principle, we need
\begin{equation} \label{kr_resolution}
  \dd (kr) = kr \, \dd \ln r + kr \, \dd \ln k \lesssim 1 \ ,
\end{equation}
so that the variations are small over one period of the trigonometric functions. The strictest limit comes from the largest values of $k$ and $r$: to get accurate results up to a given $r$, the resolutions must obey
\begin{equation} \label{resolution_limits}
  \dd \ln k \lesssim \frac{1}{\kmax r} \ , \qquad \dd \ln r \lesssim \frac{1}{\kmax r} \ ,
\end{equation}
where $\kmax$ is the maximum relevant $k$. For full resolution, we should take $\kmax = \kend$ from \re{zeta_2}, but for practical estimates, we can use $\kmax = \kpeak$, since the contribution from much larger $k$-values is suppressed by the small amplitude of the power spectrum. Changing $\kend$ has negligible effect on our results, as long as $\Pzeta(\kend) \ll \Pzeta(\kpeak)$, as shown in \fig{fig:solar_kend} for the solar case.

The $k$ and $r$ resolutions in \re{resolution_limits} have different roles. If $\dd \ln r$ is not small enough, some peaks of the compaction function may be missed. This decreases the PBH abundance, but the result still gives a reliable lower limit for the abundance. If $\dd \ln k$ is not small enough, the predicted value of $\C$ at a given $r$ becomes unreliable. In this case, the result may under- or overestimate the abundance. There are two issues. First, in the sum \re{r_zeta_prime}, the phase $kr$ in the trigonometric functions becomes essentially random, so the kernel is not properly evaluated. Second, we end up assigning only one random variable $\hxi_k$ for a range of the kernel values (instead of one for each distinguishable value), which introduces unphysical correlations between scales. This may lead to compaction function profiles that vary rapidly and have large values. For this reason, $k$ resolution is more important than $r$ resolution for the convergence of our results.

To solve the first $k$ resolution issue, we modify the sum \re{r_zeta_prime} as follows. In the original integral, consider a narrow bin from $k_1$ to $k_2$, where $\ln k_2 = \ln k_1 + \dd \ln k$ so that (compare to \re{zeta})
\begin{equation} \label{r_zeta_prime_component_integral}
  r \zeta'(r) \supset \int_{k_1}^{k_2} \frac{\dd \zeta_{<k}}{\dd \ln k} \left[\cos{} ( k r ) - \frac{\sin{}(k r)}{k r}\right]  \rmd \ln k \ .
\end{equation}
We take $\dd \zeta_{<k}/\dd \ln k$ as approximately constant over the bin (this is a good approximation since $\Pzeta(k)$ is almost constant for a narrow bin, and we consider only one noise value $\hxi_k$ there), and only integrate over the (possibly quickly varying) trigonometric kernel, yielding
\begin{equation} \label{r_zeta_prime_component_integral_2}
  r \zeta'(r) \supset \frac{\dd \zeta_{<k}}{\dd \ln k} \frac{\sin{}(k r)}{k r} \Bigg|_{k_1}^{k_2} \ .
\end{equation}
Then, instead of the approximation \re{r_zeta_prime}, we get
\bea \label{r_zeta_prime_2}
    r\zeta'(r) &=& \sum_{k=\kini}^{\kend} \Bigg[ -\frac{\hxi_k}{1-\frac{\epsilon_2}{2}X_{<k}}\sqrt{\frac{\Pzeta(k)}{\rmd \ln k}}  - \frac{\epsilon_2}{4\qty(1-\frac{\epsilon_2}{2}X_{<k})^2} \Pzeta(k) \Bigg] \el
    && \times\left[\frac{\sin{}(k r e^{\rmd \ln k})}{k r e^{\rmd \ln k}} - \frac{\sin{}(k r)}{k r}\right] \ .
\eea
Even if $kr$ varies quickly inside the interval, the result \re{r_zeta_prime_2} integrates over this variation accurately. For the individual $k$ steps, if $\dd \ln k \ll 1/(kr)$, then the kernel $\sin(k r e^{\rmd \ln k})/(k r e^{\rmd \ln k}) - \sin(k r)/(k r)$ scales as $\dd \ln k$, returning the original form $[\cos(kr)-\sin(kr)/kr]\dd \ln k$. If $\dd \ln k \gtrsim 1/(kr)$ (the problematic region for convergence), then the kernel scales as $1/(kr)$, as the rapid variations are smoothed. This suppresses contributions from the problematic regime and makes $\C$ approach zero faster for large $r$, suggesting we again get a lower limit for PBH abundance.\footnote{Applied consistently to $\zeta(r)$ in \re{zeta_2}, needed for the computation of $\avC$ through \re{avC}, a similar technique would lead to sums over $\text{Ci}(kr) - \sin(kr)/kr$, where $\text{Ci}(x)$ is the cosine integral. Because evaluating $\text{Ci}(x)$ is numerically more demanding than evaluating trigonometric functions, we instead there use the original formula \re{zeta_2} for $\zeta(r)$. As $\avC$ mainly depends on $r\zeta'(r)$ instead of $\zeta(r)$, we expect the resulting inaccuracy to be small. Furthermore, the integration kernel in \re{zeta_2} suppresses the contribution to the integral for large $kr$, in contrast to \re{r_zeta_prime}.} However, this does not fix the second issue with the $k$ resolution, the problem of assigning only one noise term to a long interval. There may be intermediary scales where this issue causes our resolution to fail and the $1/(kr)$ suppression has not yet fully kicked in.

In our simulations, unless otherwise mentioned, $\dd \ln r = 10^{-2}$ and $\dd \ln k = 10^{-3}$. With $\kmax=\kpeak$, the $\dd \ln k$ limit in \re{resolution_limits} gives the maximum accessible $r$ value $10^3 \kpeak^{-1}$; the $\dd \ln r$ limit gives the corresponding maximum $r$ value $10^2\kpeak^{-1}$. The corresponding mass limit for the $k$ ($r$) resolution is $1.9\times10^{-6(-8)} M_\odot$, $1.2\times10^{9(7)} M_\odot$, and $6.7\times10^{11(9)} M_\odot$ for the asteroid, solar, and supermassive cases, respectively.\footnote{These are the $M_{\rmax}$ mass values from \re{mass} without critical scaling and taking $\Cmax=\Cth=0.4$.} In figures \ref{fig:astmass}, \ref{fig:solarmass}, and \ref{fig:supermass} the mass distribution extends to the mass value indicated by the $k$ resolution, and then starts to decrease, presumably due to the $1/(kr)$ suppression. We see that the $k$ resolution is more important for convergence, in line with the above discussion. 

\begin{figure}[t!]
  \centering
  \includegraphics{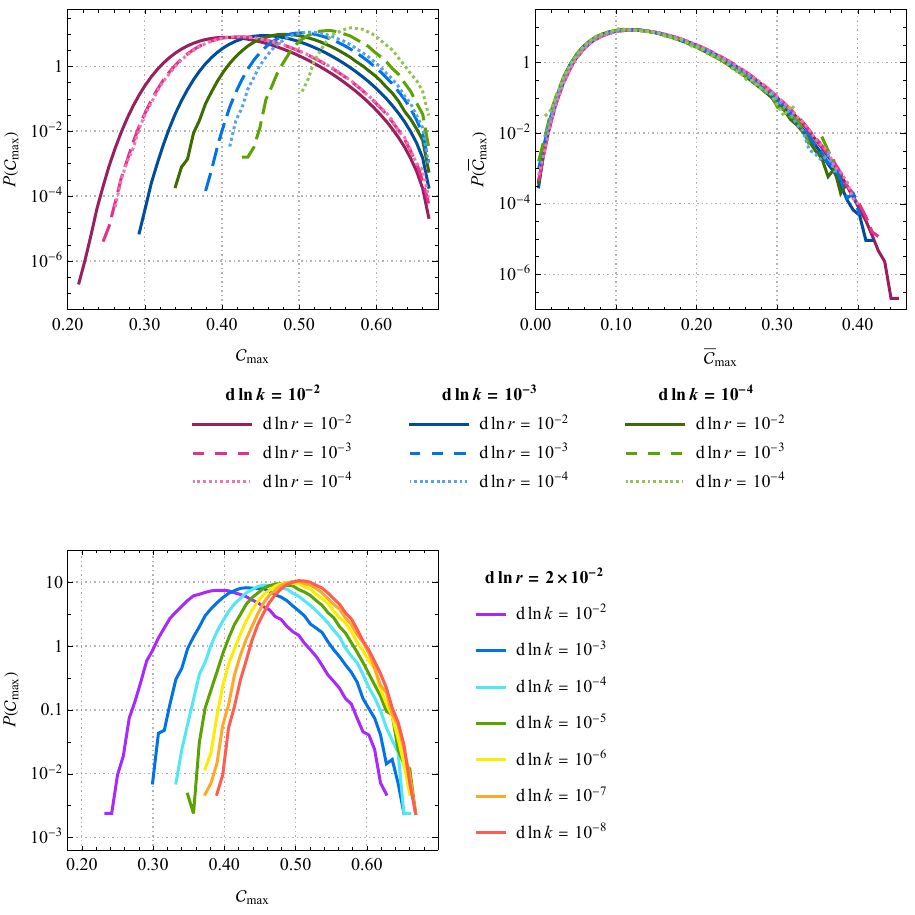}
  \caption{Convergence of the $\Cmax$ and $\avCmax$ distribution in the solar case.}
  \label{fig:convergence_c}
\end{figure}

\begin{figure}[h!]
  \centering
  \includegraphics{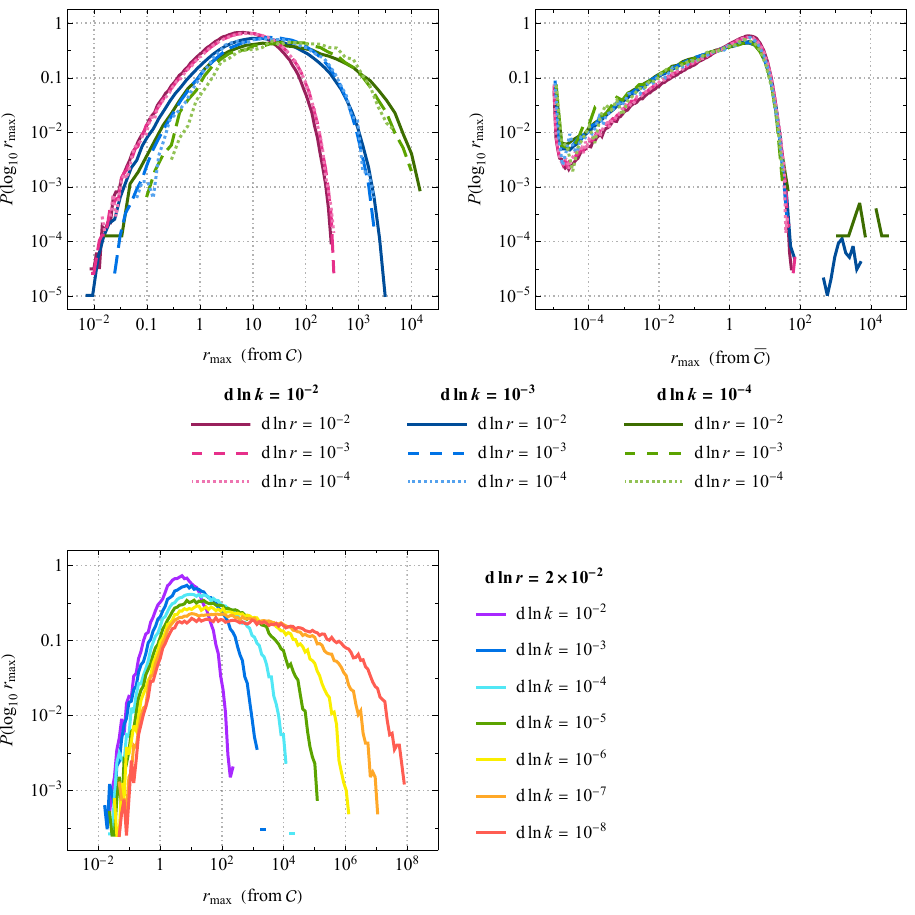}
  \caption{Convergence of the $\rmax$ distribution for $\Cmax$ and $\avCmax$ in the solar case, limited to peaks that exceed the collapse threshold.}
  \label{fig:convergence_r}
\end{figure}

Figures \ref{fig:convergence_c} and \ref{fig:convergence_r} show the full $\Cmax$ and $\avCmax$ distributions, and the $\rmax$ distribution limited to peaks that exceed the collapse threshold, for various resolutions in the solar case. Two different $k$-ranges are used in the tests. The tests that vary the $k$ resolution between $\dd \ln k = 10^{-2}$ and $\dd \ln k = 10^{-8}$ with a fixed $r$ resolution use $\kini$ and $\kend$ such that $\Pzeta(\kini)$ and $\Pzeta(\kend)$ are equal to $10^{-2}$ of the maximum value $\Pzeta(\kpeak)$. For the other tests, $\kini = k_*$ and $\kend$ is a mode that crosses the Hubble radius 10 e-folds before the end of inflation.

The $\Cmax$ distribution (and thus the PBH abundance) converges slowly. As the $k$ resolution increases, the distribution shifts towards larger values of $\Cmax$. For $\dd \ln r = 10^{-2}$, convergence in the distribution is only reached around $\dd \ln k \sim 10^{-8}$, although the abundance $\beta$ saturates to $\sim 1$ around $\dd \ln k \sim 10^{-4}$. At the same time, the corresponding distribution of $\rmax$ (which gives the PBH mass) moves up and widens. Tighter $k$ resolutions reveal more peaks of $\C$  at larger $\rmax$, which push the regime of $1/(kr)$ suppression further. There is no indication of convergence for the $\rmax$ distribution even at $\dd \ln k \sim 10^{-8}$. We have better convergence in the $r$ resolution: the $\rmax$ distribution computed from $\C$ is insensitive to it, although it still affects the $\C$ distribution somewhat.

Convergence is better for $\avC$: both the $\avC$ distribution and the corresponding $\rmax$ distribution are rather insensitive to the resolution. Presumably the averaging procedure \re{avC} washes out the high-$r$ peaks for $\avC$.

The asteroid and supermassive cases show similar behaviour, but as $\b$ does not reach unity, its value changes more with resolution. With $\dd \ln r = 10^{-2}$, $\b$ grows with improving resolution $\dd \ln k$, but the relative growth decreases with better resolution. In the asteroid case, the relative change in $\b$ (with $\Cth=0.4$) is 69\% when going from $\dd \ln k=10^{-2}$ to $\dd \ln k=10^{-3}$, and 21\% when going from $\dd \ln k=10^{-5}$ to $\dd \ln k=10^{-6}$.  The order of magnitude is stable from $\dd \ln k=10^{-2}$ to $\dd \ln k=10^{-6}$. The behaviour in the supermassive case behaviour is similar. The relative error falls slightly more slowly than in the asteroid case, but the order of magnitude does not change from $\dd \ln k=10^{-3}$ to $\dd \ln k=10^{-7}$. In both cases, we expect our results to give a good estimate for the order of magnitude of the PBH abundance, even if the prefactor is not reliable.

\bibliographystyle{JHEP}
\bibliography{coll}

\end{document}